\PassOptionsToPackage{dvipsnames}{xcolor}
\documentclass[review]{elsarticle}
\usepackage{graphicx}
\usepackage{subfig}
\usepackage{ulem}
\usepackage{amsmath}
\usepackage{mathtools}
\usepackage{xcolor}
\usepackage{hyperref}
\usepackage[capitalise]{cleveref}
\usepackage[top=1.1in, bottom=1.1in, left=1.0in, right=1.0in]{geometry}
\usepackage{lineno,accsupp}
\usepackage{bm}
\usepackage{booktabs}
\usepackage{graphics}
\usepackage{multirow}
\usepackage{booktabs}
\usepackage{cases}
\usepackage{tabularx}
\usepackage{amsfonts}
\usepackage{makecell}

\makeatletter
\let\@submitted\@empty
\let\@date\@empty
\def\ps@pprintTitle{%
  \let\@oddhead\@empty
  \let\@evenhead\@empty
  \let\@oddfoot\@empty
  \let\@evenfoot\@empty
}
\makeatother

\modulolinenumbers[1]

\biboptions{numbers,sort&compress}
\begin{document}
\title{Mitigating ray effects in rarefied flow simulations using an ensemble-of-subproblems strategy with stochastic discrete velocities}
\author[add1]{Shuyang Zhang}
\author[add2]{Weidong Li\corref{cor1}}
\ead{lwd_1982.4.8@163.com}
\author[add2]{Ming Fang}
\author[add1,add3]{Zhaoli Guo\corref{cor1}}
\ead{zlguo@mail.hust.edu.cn}
\cortext[cor1]{Corresponding author}
  \address[add1]{State Key Laboratory of Coal Combustion, School of Energy and Power Engineering, Huazhong University of Science and Technology, Wuhan 430074, China}
  \address[add2]{National Key Laboratory of Aerospace Physics in Fluids, Mianyang 621000, China}
  \address[add3]{Institute of Interdisciplinary Research for Mathematics and Applied Science, Huazhong University of Science and Technology, Wuhan 430074, China}

\begin{abstract}
  In this work, a ensemble-of-subproblems strategy with stochastic discrete velocities is extended to deterministic methods for mitigating ray effects in rarefied flow simulations.
  The strategy involves performing multiple independent subproblems, each using a small set of randomly sampled velocity points, and then averaging their solutions to obtain the final result.
  The core idea is to ensure that the distribution function at any velocity can contribute to the final result, approximating highly refined velocity-space resolution without increasing the memory requirement in any single subproblem.
  We incorporate this strategy within the DUGKS framework, and the resulting method is denoted as SDV-DUGKS.
  To evaluate the performance of the proposed method, we compare SDV-DUGKS with the original DUGKS on several test cases: (a) the Sod shock tube problem, (b) the one-dimensional Riemann problem, (c) the two-dimensional lid-driven cavity flow, and (d) the two-dimensional Riemann problem.
  The results show that, in the collisionless limit $\mathrm{Kn} \to \infty$: (1) for one-dimensional compressible flows, SDV-DUGKS reduces memory usage by approximately $2/3$ compared with that of the original DUGKS while achieving good agreement; (2) for two-dimensional compressible flows, SDV-DUGKS requires one to two orders of magnitude less memory than the original DUGKS while achieving good agreement.
  Based on these results, it can be concluded that the proposed method serves as a reliable and effective tool for mitigating ray effects in rarefied flow simulations.
\end{abstract}
\begin{keyword}
Rarefied gas flow  \sep  Ray effect  \sep  Discrete unified gas kinetic scheme \sep  Ensemble-of-subproblems strategy \sep  Stochastic discrete velocity method
\end{keyword}
\maketitle

\section{Introduction}\label{sec1}

Ray effects, initially referring to nonphysical oscillations in discrete ordinates ($S_N$) solutions of the radiation transfer equations, arising from angular discretization that breaks rotational invariance~\cite{lathrop1968ray,modest2021radiative,morel2003analysis}.
Over the years, various mitigation strategies have been developed, including fictitious-source methods~\cite{lathrop1971remedies,jung1972discrete,reed1972spherical,miller1977ray}, regional angular refinement (RAR) techniques~\cite{longoni2001investigation,stone2007adaptive,jarrell2011discrete,lau2017discrete}, and stochastic methods based on multiple random rotations of the reference frame~\cite{tencer2016ray} or random angular samplings~\cite{li2024random}.

In rarefied gas flow simulations, deterministic methods solve the Boltzmann equation by discretizing the continuous velocity space, e.g., the discrete velocity method (DVM)~\cite{platkowski1988discrete}, the unified gas kinetic scheme (UGKS)~\cite{xu2010unified,xu2014direct}, and the discrete unified gas kinetic scheme (DUGKS)~\cite{guo2013discrete,guo2015discrete}.
Similarly, restricting the velocity space to a set of fixed discrete points inevitably neglects potentially important information in the moment calculation, which also gives rise to ray effects in the numerical results, especially at large Knudsen numbers.
For example, in the Sod shock tube problem in the free molecular regime, the macroscopic profiles obtained with DVM~\cite{brull2014local} and UGKS~\cite{zhu2020ray} exhibit several plateaux;
in the $2\mathrm{D}$ lid-driven cavity flow with large Knudsen numbers, wavy temperature contours are clearly observed in the solutions of DVM~\cite{ho2019comparative}, UGKS~\cite{zhu2016implicit}, and DUGKS~\cite{zhu2016discrete}.

Recently, several studies have investigated the characteristics of ray effects in rarefied gas flow simulations.
Sekaran et al. observed that a highly refined physical grid necessitates a sufficiently high velocity grid resolution to accurately resolve small field variations in higher-order moments~\cite{sekaran2018analysis}.
Aoki et al. mathematically derived that the strength of ray effects decays exponentially with time in the presence of particle collisions~\cite{aoki2001note}.
Ho et al. reported that ray effects increase with Knudsen number and are closely related to the compatibility of resolutions between the velocity space grid and the physical space grid~\cite{ho2019comparative}.
Zhu et al. further noted that the discretization of the velocity space introduces discontinuities in the distribution function, which cause sudden changes in macroscopic variables; with the same set of discrete velocity points, different numerical integration methods produce similar ray effects~\cite{zhu2020ray}.

Consequently, the most straightforward approach to mitigating ray effects is to use a higher-resolution velocity space grid.
However, this leads to an exponential increase in computational cost with respect to the dimensionality $D$ of the velocity space.
Without employing excessively refined velocity grids, mitigating ray effects under specific flow conditions has been extensively explored.
For the ray effect caused by boundary induced discontinuities, some studies~\cite{aoki2001note,naris2005driven} have proposed treating the propagation of each discontinuity analytically while solving the remaining numerically, which is impractical for simulations with complex geometries.
For low-speed rarefied flows, a velocity grid in polar coordinates has been reported to greatly mitigate ray effects compared to a Cartesian grid~\cite{ho2019comparative}, but its performance in high-speed rarefied flows remains unclear.
For compressible rarefied flows, a local discrete velocity grid has been introduced into the DVM, in which the bounds and resolution of the velocity grid are updated at each time step based on the local velocity and temperature~\cite{brull2014local}.
While this approach partially reduces ray effects, it still requires a large number of velocity points and does not satisfy the conservation laws.
As a further step, based on an systematic analysis of ray effects induced by different origins across multiple classic test cases, covering low-speed and compressible flows at various Knudsen numbers, Zhu et al. proposed three principles for designing unstructured velocity grids to mitigate ray effects~\cite{zhu2020ray}.
However, as the authors note, the optimal discretization of the velocity space requires case-by-case design.

In summary, most existing studies have, conceptually, relied on locally refining the velocity space resolution based on macroscopic flow condition.
While this can mitigate ray effects without increasing the number of velocity points, it remains problem-dependent.
Beyond specific macroscopic flow conditions, the unifying principle for mitigating ray effects is to ensure that the distribution function is adequately resolved in velocity space.
To realize this, we introduce an ensemble-of-subproblems strategy with stochastic discrete velocities, proposed in our previous work, into deterministic methods.
The proposed strategy consists of three key steps: (a) perform multiple independent simulations for the problem of interest; (b) use a small set of randomly sampled velocity points in each simulation; (c) average their results to obtain the final solution.
Through averaging based on multiple sparse samplings, this strategy aims to approximate the resolution of a finely discretized velocity space without requiring its full computational cost.
Applied to DUGKS, this method, referred to as SDV-DUGKS, has been reported as a reliable tool for rarefied flow simulations ranging from low-speed to supersonic conditions.
In the present paper, we focus on evaluating the capability of this strategy to mitigate ray effects in rarefied flow simulations.
For this purpose, we conduct a comparative study between SDV-DUGKS and the original DUGKS by performing various test cases in the collisionless limit $\mathrm{Kn} \to \infty$.

The remainder of this paper is structured as follows.
In Sec.~\ref{sec2}, the procedure of the SDV-DUGKS is presented.
In Sec.~\ref{sec3}, several numerical tests are performed to demonstrate the ability of the SDV-DUGKS to mitigate ray effects.
In Sec.~\ref{sec4}, a concise summary is provided.

\section{Numerical methods}\label{sec2}
\subsection{Discrete velocity Boltzmann equation}\label{sec21}
In this work, the widely used Shakhov model~\cite{shakhov1968generalization} is adopted,
\begin{equation}
    \frac{\partial f}{\partial t}+\bm \xi\cdot\nabla f
    =\Omega\equiv-\frac{1}{\tau}[f-f^{S}],
\label{eq:ShakhovModel}
\end{equation}
where $ f = f(\bm x, \bm \xi, \bm \eta, \bm \zeta, t) $ is the distribution function of particles in a $D$-dimensional physical space with velocity $ \bm\xi = (\xi_{1}, \dots, \xi_{D}) $ at position $ \bm x = ( x_{1}, \dots,  x_{D}) $ and time $ t $.
In this expression, $ \bm\eta \in \mathbb{R}^{3-D} $ denotes the remaining velocity components in the three-dimensional velocity space;
$\bm \zeta\in \mathbb{R}^{K}$ represents the internal degrees of freedom.
The equilibrium distribution function $f^S$ is given by
\begin{equation}
    f^S=f^{eq}\Bigg[1+(1-\mathrm{Pr})\frac{\bm c\cdot \bm q}{5pRT}\Big(\frac{c^{2}+\eta^{2}}{RT}-5\Big)\Bigg],
\label{eq:ShakhovEq}
\end{equation}
with $f^{eq}={\rho}{(2\pi R T)^{-3/2}}\exp\left[-{ \left(c^{2}+\eta^{2}+\zeta^{2}\right)}/{\left(2RT\right)}\right]$.
In the above, $ \rho $ is the gas density, $\bm{c}=\bm{\xi} - \bm{u}$ is the peculiar velocity, $ \bm{q} $ is the heat flux, $ R $ is the gas constant, and $ T $ is the temperature.
The relaxation time $ \tau $ is related to the dynamic viscosity $ \mu $ and pressure $ p $ via $\tau = \mu/{p}$.

To remove the dependence on $\bm\eta$ and $\bm\zeta$, reduced distribution functions, $g$ and $h$, are introduced to characterize the velocity field and energy field,
\begin{subequations}
    \begin{equation}
    g(\bm x,\bm\xi,t)=\int f d\bm\eta d\bm\zeta, \\
    \label{eq:Distributiong}
    \end{equation}
    \begin{equation}
    h(\bm x,\bm\xi,t)=\int (\eta^2+\zeta^2)fd\bm\eta d\bm\zeta.\\
    \label{eq:Distributionh}
    \end{equation}
    \label{eq:Distributionf}%
\end{subequations}
Correspondingly, the reduced equilibrium distribution functions, $g^S$ and $h^S$, are given by
\begin{subequations}
    \begin{equation}
    \begin{aligned}
    \quad g^S={g^{eq}}\left\{1+(1-\mathrm{Pr}){\frac{\bm c\cdot \bm q}{5p R T}}\left[\frac{c^{2}}{R T}-D-2\right]\right\},
    \label{eq:ReducedgS}
    \end{aligned}
    \end{equation}
    \begin{equation}
    \begin{aligned}
    h^S
    =(3-D+K)RT{g}^{eq}+(1-\mathrm{Pr}){\frac{\bm c\cdot \bm q}{5p R T}}\left[\left(\frac{c^{2}}{RT}-D\right)(3-D+K)-2K\right]RT{g}^{eq},
    \label{eq:ReducedhS}
    \end{aligned}
    \end{equation}
    \label{eq:ReducedfS}%
\end{subequations}
with ${g^{eq}}={\rho}{(2\pi R T)}^{-D/2}\exp{\left[-{c^{2}}/{(2RT)}\right]}$.

A key step in numerically solving the Boltzmann equation within a deterministic framework is to restrict the distribution functions $g$ and $h$ to a discrete set of velocities $\{\bm{\xi}_1, \dots, \bm{\xi}_N\} \subset \mathbb{R}^D$. This discrete representation of the velocity space transforms the Boltzmann equation into the discrete velocity Boltzmann equation (DVBE):
\begin{equation}
\frac{\partial \phi_{\alpha}}{\partial t} + \bm{\xi}_{\alpha} \cdot \nabla \phi_{\alpha} = \Omega(\phi_{\alpha}) \equiv -\frac{1}{\tau} [\phi_{\alpha} - \phi_{\alpha}^S],  \quad \alpha = 1,\dots,N,
\label{eq:phiEquation}
\end{equation}
where $\phi = g$ or $h$, and $\phi_{\alpha}(\bm{x},t) \coloneqq \phi(\bm{x},\bm{\xi}_{\alpha},t)$ and $\phi_{\alpha}^S(\bm{x},t) \coloneqq \phi^S(\bm{x},\bm{\xi}_{\alpha},t)$ represent the corresponding discrete distribution function and equilibrium distribution function at discrete velocity $\bm{\xi}_{\alpha}$, respectively.
The conserved variables $ \bm W=(\rho, \rho \bm u, \rho E)^T $ are obtained via numerical quadrature over the discrete velocity set,
\begin{equation}
    \rho=\sum_{\alpha=1}^{N}\omega_{\alpha}{g}_{\alpha},\quad \rho \bm u=\sum_{\alpha=1}^{N}\omega_{\alpha}\bm{\xi}_{\alpha}{g}_{\alpha},
    \quad\rho E=\frac{1}{2}\sum_{\alpha=1}^{N}\omega_{\alpha}\left({\xi}^2_{\alpha}{g}_{\alpha}+{h}_{\alpha}\right), \\
\label{eq:gCMoments}
\end{equation}
where $\omega_{\alpha}$ denotes the quadrature weight for $\bm{\xi}_{\alpha}$.
$\rho E={\textstyle{\frac{1}{2}}}\rho u^{2}+\rho c_{V}T$ represents the total energy, where $c_{V}$ denotes the specific heat capacity at constant volume.
Moreover, the heat flux $\bm{q}$ is given by
\begin{equation}
    \bm{q}=\frac{1}{2}\sum_{\alpha=1}^{N}\omega_{\alpha}\bm{c}_{\alpha}\left(c_{\alpha}^{2}{g}_{\alpha}+{h}_{\alpha}\right), \\
\label{eq:gq}
\end{equation}
with $\bm{c}_{\alpha}=\bm{\xi}_{\alpha}-\bm{u}$.
For numerical accuracy and physical consistency, the numerical moments of the collision term must satisfy the conservation laws of mass, momentum, and total energy, i.e.,
\begin{equation}
    \sum_{\alpha=1}^{N}\omega_{\alpha}\Omega_{{g}_{\alpha}}=0, \quad \sum_{\alpha=1}^{N}\omega_{\alpha}\bm{\xi}_{\alpha}\Omega_{{g}_{\alpha}}=0, \quad
    \sum_{\alpha=1}^{N}\omega_{\alpha}\left({\xi}_{\alpha}^2\Omega_{{g}_{\alpha}}+\Omega_{{h}_{\alpha}}\right)=0. \\
\label{eq:conservation_laws}
\end{equation}

\subsection{Discrete unified gas kinetic scheme}\label{sec22}
Although the present work aims to mitigate ray effects, the DUGKS is introduced here as the representative method, since the proposed modifications are built upon it.
Integrating Eq.~\eqref{eq:phiEquation} over the control volume $ V_j $ centered at $ \bm x_j $ from $ t_n $ to $ t_{n+1} $ gives
\begin{equation}
    \phi_{j,\alpha}^{n+1}-\phi_{j,\alpha}^{n}+\frac{\Delta t}{|V_{j}|}\bm F_{j,\alpha}^{n+1/2}
    =\frac{\Delta t}{2}\left(\Omega_{j,\alpha}^{n+1}+\Omega_{j,\alpha}^{n}\right),
\label{eq:phievelotion}
\end{equation}
where the midpoint rule and the trapezoidal rule are applied to the convection and collision terms, respectively.
The microflux across the cell interface, denoted as $ \bm{F}_{j,\alpha}^{n+1/2} $, is defined by
\begin{equation}
    \bm{F}_{j,\alpha}^{n+1/2}=\int_{\partial V_j}\,(\bm{\xi}_{\alpha}\cdot \bm n){f}_{\alpha}^{n+1/2}\, d\bm{S},
\label{eq:microfluxF}
\end{equation}
where $\partial V_j$ representing the cell surface and $\bm n$ the outward unit normal vector.

To remove the implicitness of the collision term, DUGKS introduces the auxiliary distributions defined as
\begin{equation}
    \tilde{\phi}=\phi-{\frac{\Delta t}{2}}\Omega,\quad \tilde{\phi}^{+}=\phi+{\frac{\Delta t}{2}}\Omega.
\label{eq:auxiliaryphitilde}
\end{equation}
Eq.~\eqref{eq:phievelotion} can then be rewritten as
\begin{equation}
    \tilde{\phi}_{j,\alpha}^{n+1}=\tilde{\phi}_{j,\alpha}^{+,n}-\frac{\Delta t}{|V_{j}|}\bm{F}_{j,\alpha}^{n+1/2}.
\label{eq:tildephievelution}
\end{equation}
From the conservation properties of the collision term in Eq.~\eqref{eq:conservation_laws}, the conserved variables are computed as
\begin{equation}
    \rho=\sum_{\alpha=1}^{N}\omega_{\alpha}\tilde{g}_{\alpha},\quad
    \rho \bm u=\sum_{\alpha=1}^{N}\omega_{\alpha}\bm\xi_{\alpha}\tilde{g}_{\alpha},\quad
    \rho E=\frac{1}{2}\sum_{\alpha=1}^{N}\omega_{\alpha}\left(\xi_{\alpha}^{2}\tilde{g}_{\alpha}+\tilde{h}_{\alpha}\right),
\label{eq:gTlideCMoments}
\end{equation}
and the heat flux $\bm q$ is given by
\begin{equation}
    \bm q=\frac{2\tau}{2\tau+\Delta t\mathrm{Pr}}\tilde{\bm q},\,\quad \mathrm{with}\ \quad
    \tilde{\bm q}=\frac{1}{2}\sum_{\alpha=1}^{N}\omega_{\alpha}\bm{c}_{\alpha}\left(c_{\alpha}^{2}\tilde{g}_{\alpha}+\tilde{h}_{\alpha}\right).
\label{eq:gTlideq}
\end{equation}
Given the microflux $ \bm F^{n+1/2} $, the distribution function $\tilde{\phi}$ can be explicitly updated via Eq.~\eqref{eq:tildephievelution}.

To evaluate $\bm F^{n+1/2}$, the distribution function $\phi^{n+1/2}$ at the cell interface center is needed.
Integrating Eq.~\eqref{eq:phiEquation} along the characteristic line over a half time step $s=\Delta t/2$ and applying the trapezoidal rule to the collision term yields
\begin{equation}
    \phi_{\alpha}\left(\bm x_{b},t_{n}+s\right)-\phi_{\alpha}\left(\bm x_{b}-\bm{\xi}_{\alpha} s,t_{n}\right)
    =\frac{s}{2}\left[\Omega_{\alpha}\left(\bm x_{b},t_{n}+s\right)
    +\Omega_{\alpha}\left(\bm x_{b}-\bm{\xi}_{\alpha} s,t_{n}\right)\right],
\label{eq:halfphievelotion}
\end{equation}
where $\bm x_b$ denotes the interface center of cell $j$.
Similar to Eq.~\eqref{eq:auxiliaryphitilde}, two auxiliary distribution functions are introduced,
\begin{equation}
    \bar{\phi}=\phi-{\frac{s}{2}}\Omega,\quad \bar{\phi}^{+}=\phi+{\frac{s}{2}}\Omega.
\label{eq:auxiliaryphibar}
\end{equation}
Eq.~\eqref{eq:halfphievelotion} can then be rewritten as
\begin{equation}
    \bar{\phi}_{\alpha}(\bm x_{b},t_{n+1/2})=\bar{\phi}_{\alpha}^{+}(\bm x_{b}-\bm{\xi}_{\alpha} s,t_{n}).
\label{eq:barphievelotion}
\end{equation}
The term $ \bar{\phi}_{\alpha}^{+}(\bm x_{b}-\bm{\xi}_{\alpha} s,t_{n}) $ is obtained via linear reconstruction:
\begin{equation}
    \bar{\phi}_{\alpha}^{+}(\bm x_{b}-\bm{\xi}_{\alpha} s,t_{n})=\bar{\phi}_{\alpha}^{+}(\bm x_{j},t_{n})
    +(\bm x_{b}-\bm x_{j}-\bm{\xi}_{\alpha} s)\cdot\bm\sigma_{j},
\label{eq:phiBarinterfaceGot}
\end{equation}
where $\bm\sigma_{j}$ denotes the slope of $\bar{\phi}^{+}$ in cell $j$.
With $\bar{\phi}_{\alpha}(\bm x_{b},t_{n+1/2})$ known, the macroscopic variables at time $t_{n+1/2}$ are computed as
\begin{equation}
    \rho=\sum_{\alpha=1}^{N}\omega_{\alpha}\bar{g}_{\alpha},\quad
    \rho \bm u=\sum_{\alpha=1}^{N}\omega_{\alpha}\bm\xi_{\alpha}\bar{g}_{\alpha},\quad
    \rho E=\frac{1}{2}\sum_{\alpha=1}^{N}\omega_{\alpha}(\xi_{\alpha}^{2}\bar{g}_{\alpha}+\bar{h}_{\alpha}),
\label{eq:gBarCMoments}
\end{equation}
and
\begin{equation}
    \bm q=\frac{2\tau}{2\tau+s\mathrm{Pr}}\bar{\bm q},\,\quad \mathrm{with}\ \quad
    \bar{\bm q}=\frac{1}{2}\sum_{\alpha=1}^{N}\omega_{\alpha}\bm{c}_{\alpha}\left(c_{\alpha}^{2}\bar{g}_{\alpha}+\bar{h}_{\alpha}\right).
\label{eq:gBarq}
\end{equation}
With these quantities, the Shakhov equilibrium distribution function $\phi_{\alpha}^{S}(\bm x_{b},t_{n}+s)$ can be evaluated from Eq.~\eqref{eq:ReducedfS}.
Finally, the original distribution function at the cell interface is calculated via Eq.~\eqref{eq:auxiliaryphibar},
\begin{equation}
    \phi_{\alpha}\left(\bm x_b,t_{n+1/2}\right)=\frac{2\tau}{2\tau+s}\bar{\phi}_{\alpha}\left(\bm x_b,t_n+s\right)
    +\frac{s}{2\tau+s}\phi_{\alpha}^S\left(\bm x_b,t_n+s\right).
\label{eq:phiBarShakhov}
\end{equation}

\subsection{Discrete unified gas kinetic scheme using an ensemble-of-subproblems strategy with stochastic discrete velocities (SDV-DUGKS)}\label{sec23}
As noted in Ref.~\cite{zhu2020ray,ho2019comparative}, the ray effect originates from a mismatch between the resolution in physical space and that in velocity space.
Specifically, for a given physical space mesh, the ray effect becomes pronounced when the velocity space discretization fails to resolve sharp gradients of the distribution function.
In deterministic methods, directly refining the velocity space mesh can effectively alleviate this phenomenon, but the computational cost grows exponentially with the velocity space dimension $D$.

To mitigate the ray effect while controlling the computational cost, this study incorporates an ensemble-of-subproblems strategy with stochastic discrete velocities within the DUGKS framework, yielding a scheme denoted as SDV-DUGKS.
The strategy involves performing multiple independent simulations, each using a small set of randomly sampled velocity points, and then averaging the solutions to obtain the final flow field.
Each random set defines a subproblem, which is solved following the original DUGKS procedure
However, macroscopic moments are evaluated via Monte Carlo integration~\cite{robert1999monte}, and a correction step is introduced to enforce conservation laws~\cite{zhang2026microscopically}, as detailed in the following subsections. 
By randomizing the velocity sets across simulations, the distribution function at any velocity can contribute to the averaged solution. 
In this way, the proposed approach approximates a finer velocity-space resolution without increasing velocity points in any single subproblem.
In addition, this strategy can be easily extended to other deterministic methods.

\subsubsection{Stratified sampling in velocity space}\label{sec231}
In contrast to deterministic methods, where discrete velocities are determined by a fixed velocity space mesh, the present method randomly samples discrete velocities from the velocity space domain.
This ensures that the distribution function at any velocity has the opportunity to contribute to the final result.
In this work, stratified sampling~\cite{caflisch1998Monte,lorek2025variance} is employed.
Compared with traditional random sampling techniques, it improves sampling efficiency and enhances the accuracy of Monte Carlo integration~\cite{lorek2025variance}.

Specifically, we define the three-dimensional parameter space as a truncated, cubic domain $V\subset \mathbb{R}^{3}$ in velocity space.
Stratified sampling is then performed in two steps:
\begin{enumerate}[(a)]
  \item
    stratify the continuous parameter space $V$ into $N$ disjoint, equiprobable cells $V_\alpha$ $(\alpha = 1,\dots,N)$, each of volume $\mathcal{V}_\alpha$;
  \item
    randomly sample one velocity point $\bm{\xi}_\alpha$ from each cell $V_\alpha$ according to a uniform distribution.
\end{enumerate}
The stratification can be implemented using either a structured grid or a problem-specific unstructured grid.
In this work, stratification based on a structured velocity space grid yields satisfactory performance.

\subsubsection{Computation of macroscopic quantities}\label{sec232}
As in the original DUGKS, the computation of macroscopic quantities at cell centers and interfaces requires moment evaluation.
With the integration nodes $\{\bm{\xi}_\alpha\}$ obtained, the macroscopic quantities are evaluated via Monte Carlo integration over the velocity domain as follows:
\begin{equation}
    \begin{pmatrix}\rho
 \\\rho\bm{u}
 \\2\rho{E}
 \\2\bm{q}
\end{pmatrix}=\sum^{N}_{\alpha=1}\mathcal{V}_{\alpha}[\begin{pmatrix}1
                           \\\bm\xi_\alpha
                           \\\xi_\alpha^2
                           \\\bm{c}_\alpha(c_\alpha)^2
                           \end{pmatrix}g_{\alpha}
                           +\begin{pmatrix}0
                           \\\bm{0}
                           \\1
                           \\\bm{c}_\alpha
                           \end{pmatrix}h_{\alpha}].
\label{eq:sum_macro_moments_stoca}
\end{equation}
To achieve higher effective resolution without increasing memory requirements, we perform $M$ independent simulations, each using $N$ velocity samples for Monte Carlo integration. The final result is then obtained by averaging the $M$ solutions:
\begin{equation}
    \begin{pmatrix}\rho
 \\\rho\bm{u}
 \\\rho{E}
 \\\bm{q}
\end{pmatrix}=\frac{1}{M}\sum_{m=1}^{M}\begin{pmatrix}\rho_{(m)}
 \\\rho_{(m)}\bm{u}_{(m)}
 \\\rho_{(m)}{E}_{(m)}
 \\\bm{q}_{(m)}
\end{pmatrix}, \\
\label{eq:average_M}
\end{equation}
where the subscript $m$ denotes the realization index.

\subsubsection{The microscopically conservation-enforced correction}\label{sec233}
For numerical accuracy and physical consistency, the conservative constraints in Eq.~\eqref{eq:conservation_laws} must be satisfied.
In practice, however, using sampled velocity points for Monte Carlo integration introduces an unphysical numerical source term $({1}/{\tau})\bm{R}$ into the RHS of the macroscopic conservation equations, with $\bm{R}$ given by
\begin{equation}
    \bm{R}=\tau\sum^{N}_{\alpha=1}\mathcal{V}_{\alpha}(\begin{pmatrix}1
                           \\\bm\xi_\alpha
                           \\{\xi_\alpha}^2
                           \end{pmatrix}\Omega_{g,\alpha}
                           +\begin{pmatrix}0
                           \\\bm{0}
                           \\1
                           \end{pmatrix}\Omega_{h,\alpha})
          =\sum^{N}_{\alpha=1}\mathcal{V}_{\alpha}\begin{pmatrix}{g}^S_\alpha
                           \\\bm\xi_{\alpha}{g}^S_\alpha
                           \\(\xi_{\alpha}^2{g}^S_\alpha+{h}^S_\alpha)
                           \end{pmatrix}
                          -\begin{pmatrix}\rho
                          \\\rho\bm{u}
                          \\2\rho{E}
                          \end{pmatrix}.
\label{eq:numerical_error}
\end{equation}
This term accumulates over successive iterations and may adversely affect numerical stability~\cite{titarev2007conservative,zhang2026microscopically}.
With limited integration nodes, conservation enforcement is a fundamental requirement for securing accurate simulations.

Accordingly, the microscopically conservation-enforced DUGKS (MicroC-DUGKS) from our prior work~\cite{zhang2026microscopically} is adopted.
This method introduces correction steps into the original DUGKS procedure, performed after initializing the macroscopic flow field and evaluating macroscopic quantities at cell centers and interfaces.
These steps re-evaluate the Shakhov equilibrium distribution to ensure the conservation of mass, momentum, and total energy.
To this end, a variable $\bm{A}^* = (\rho^*, \bm{u}^*, T^*, \bm{q}^*)^T$ (perturbed state of $\bm{A}=(\rho, \bm{u}, T, \bm{q})^T$) is determined such that the unphysical numerical source term vanishes, i.e.,
\begin{equation}
    \bm{R}(\bm{A}^*)=\sum_{\alpha=1}^{N}\mathcal{V}_{\alpha}\begin{pmatrix}{g}^{S}_{\alpha}(\bm{A}^*)
                                     \\\bm\xi_{\alpha}{g}^{S}_{\alpha}(\bm{A}^*)
                                     \\\left[\xi_{\alpha}^2{g}^{S}_{\alpha}(\bm{A}^*)+{h}^{S}_{\alpha}(\bm{A}^*)\right]
                                     \end{pmatrix}
                          -\begin{pmatrix}\rho
                          \\\rho\bm{u}
                          \\2\rho{E}
                          \end{pmatrix}\equiv\bm{0}. \\
\label{eq:numerical_error_null}
\end{equation}
Beyond the conservation constraints, an additional constraint on the heat flux can also be incorporated to ensure accurate heat flux evaluation:
\begin{equation}
    \sum_{\alpha=1}^{N}\mathcal{V}_{\alpha}\left\{\bm{c}_{\alpha}^*\left[({c}_{\alpha}^*)^{2}{g}^{S}_{\alpha}(\bm{A}^*)+{h}^{S}_{\alpha}(\bm{A}^*)\right]-
    \bm{c}_{\alpha}({c}_{\alpha}^2{g}_{\alpha}+{h}_{\alpha})\right\}=-2\mathrm{Pr}\bm{q}, \\
\label{eq:heat flux_corrected}
\end{equation}
where $\bm{c}_\alpha^* = \bm{\xi}_\alpha - \bm{u}^*$.

Together, the conservation and heat-flux constraints form a nonlinear system for $\bm{A}^*$,
\begin{equation}
\bm{R}^{\prime}(\bm{A}^*)=
                           \sum_{\alpha=1}^{N}\mathcal{V}_{\alpha}(\begin{pmatrix}1
                           \\\bm\xi_{\alpha}
                           \\\xi_{\alpha}^2
                           \\\bm{c}_{\alpha}^*(c_{\alpha}^*)^2
                           \end{pmatrix}{g}^{S}_{\alpha}(\bm{A}^*)
                           +\begin{pmatrix}0
                           \\\bm{0}
                           \\1
                           \\\bm{c}_{\alpha}^*
                           \end{pmatrix}{h}^{S}_{\alpha}(\bm{A}^*))
                           -\begin{pmatrix}
 \rho\\
 \rho\bm{u}\\
 2\rho{E}\\
2(1-\mathrm{Pr})\bm{q}
\end{pmatrix}\equiv\bm{0}, \\
\label{eq:conservation_heatflux_corrected}
\end{equation}
which can be solved using Newton's method with the known macroscopic variables $\bm{A}$ as the suitable initial guess $\bm{A}^0$.
Notably, the resulting $\bm{A}^*$ is used only to recompute the equilibrium distributions $g^{S}(\bm{A}^*)$ and $h^{S}(\bm{A}^*)$ for the correction; the original macroscopic variables $\bm{A}$ at cell centers and interfaces remain unchanged.
In practice, the Newton process converges rapidly, typically within one or two iterations~\cite{titarev2007conservative,huang2011conservative}.

\subsection{Algorithm}\label{sec24}
\begin{figure}[!ht]
\centering
  \subfloat{\includegraphics[width=0.9\textwidth]{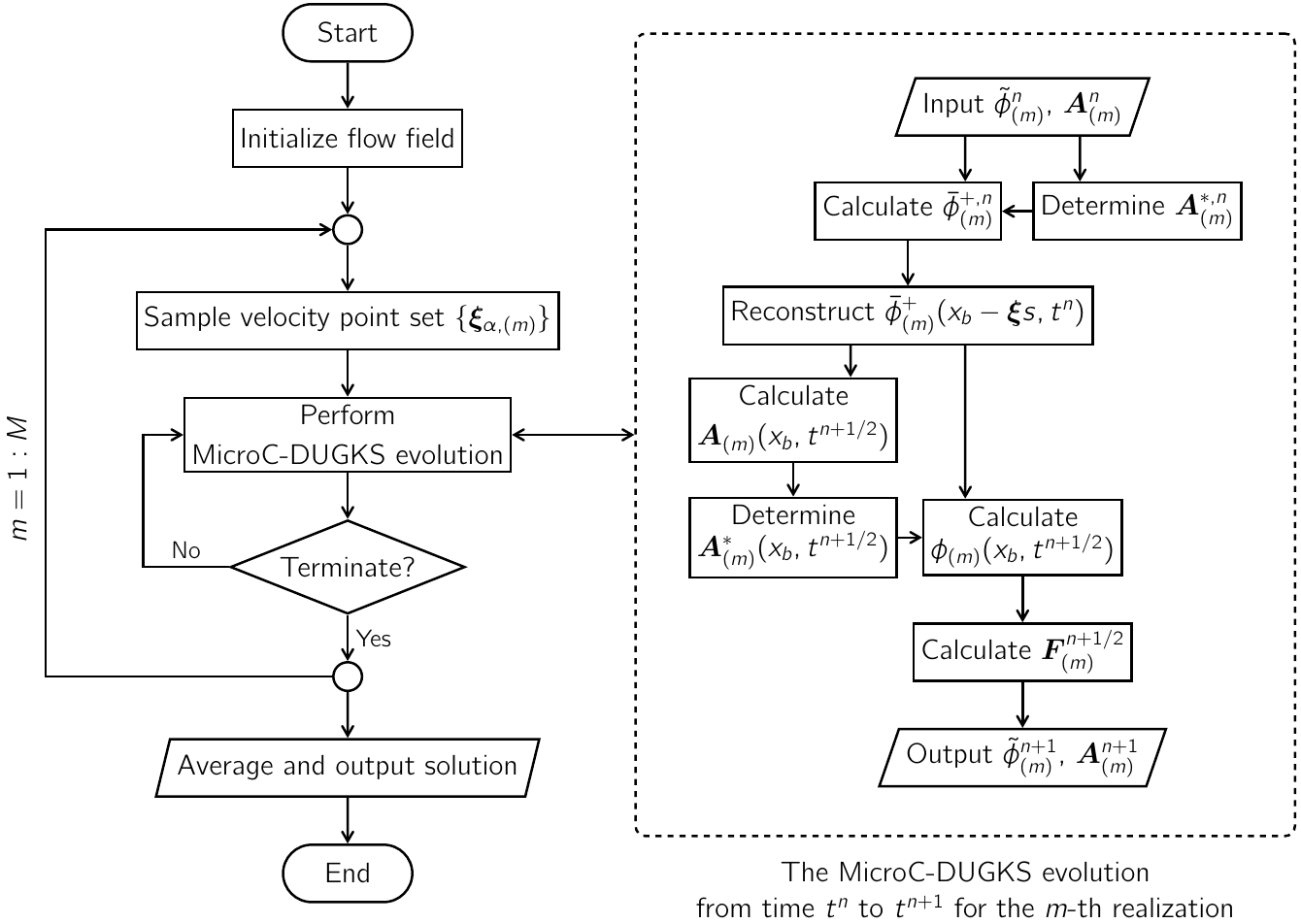}}~~
 \caption{Flow chart of the SDV-DUGKS.}
\label{Flow_chart_241}
\end{figure}

For clarity, the procedure of SDV-DUGKS is illustrated in Fig.~\ref{Flow_chart_241} and summarized as follows:
\begin{enumerate}
\item Pre-processing step
    \begin{enumerate}
        \item Set the number of independent realizations $M$.
        \item Initialize the macroscopic flow field.
        \item For each realization, sample a velocity point set $\{\bm{\xi}_\alpha\}$ and initialize its corresponding distribution functions $\{\tilde{\phi}_{\alpha}\}$ at $t_0$.
    \end{enumerate}
\item Evolution procedure of the MicroC-DUGKS from $t_n$ to $t_{n+1}$ for each realization
    \begin{enumerate}
        \item Determine the corrected state $\bm{A}^*$ at each cell center and time $t_{n}$ by solving Eq.~\eqref{eq:conservation_heatflux_corrected}.
        \item Compute the auxiliary distribution $\bar{\phi}^+_{\alpha}$ at each cell center and $t_n$ from $\tilde{\phi}_{\alpha}$ and $\phi^{S}_{\alpha}(\bm{A}^*)$ according to Eqs.~\eqref{eq:auxiliaryphitilde} and~\eqref{eq:auxiliaryphibar}.
        \item Reconstruct $\bar{\phi}^+_{\alpha}$ at $(x_b-\bm{\xi}_{\alpha}s)$ according to Eq.~\eqref{eq:phiBarinterfaceGot}.
        \item Obtain the distribution function $\bar{\phi}_{\alpha}$ at $x_b$ and $t_{n+1/2}$ according to Eq.~\eqref{eq:barphievelotion}.
        \item Evaluate the conservative macroscopic variables $\bm{W}(x_b,t_{n+1/2})$ and heat flux $\bm{q}(x_b,t_{n+1/2})$ from $\bar{\phi}_{\alpha}$ according to Eqs.~\eqref{eq:gBarCMoments}, ~\eqref{eq:gBarq} and~\eqref{eq:sum_macro_moments_stoca}.
        \item Determine the corrected state $\bm{A}^*$ at each cell interface and time $t_{n+1/2}$ by solving Eqs.~\eqref{eq:conservation_heatflux_corrected}.
        \item Calculate the original distribution function $\phi_{\alpha}$ at $x_b$ and $t_{n+1/2}$ from $\bar{\phi}_{\alpha}$ and $\phi^{S}_{\alpha}(\bm{A}^*)$ according to Eq.~\eqref{eq:phiBarShakhov}.
        \item Calculate the microflux $\bm{F}_{\alpha}^{n+1/2}$ across the cell interface from $\phi_{\alpha}^{n+1/2}$ according to Eq.~\eqref{eq:microfluxF}.
        \item Update $\tilde{\phi}_{\alpha}$ at each cell center and $t_{n+1}$ from $\bm{F}_{\alpha}^{n+1/2}$ according to Eqs.~\eqref{eq:auxiliaryphitilde} and~\eqref{eq:tildephievelution}.
    \end{enumerate}
\item Post-processing step
    \begin{enumerate}
        \item Upon termination of each realization, evaluate the macroscopic variables at each cell center according to Eqs.~\eqref{eq:gTlideCMoments} and~\eqref{eq:gTlideq}.
        \item Average the results over all $M$ realizations and output the final macroscopic flow field according to Eq.~\eqref{eq:average_M}.
    \end{enumerate}
\end{enumerate}

\section{Numerical results and discussions}\label{sec3}
This section validates the effectiveness of SDV-DUGKS in mitigating ray effects through several test problems: the Sod shock tube problem, $1\mathrm{D}$ Riemann problem, $2\mathrm{D}$ lid-driven cavity flow, and $2\mathrm{D}$ Riemann problem. The gas constant and Prandtl number are set to $R=0.5$ and $\mathrm{Pr}=2/3$, respectively.
All simulations are performed on an Intel Xeon Gold 6348 CPU @ 2.60 GHz processor.

In the simulations, the distribution function is initialized as the Maxwellian distribution for the initial conditions.
The solution is advanced in time until the termination condition is satisfied.

\subsection{Sod shock tube problem}\label{sec31}
\begin{figure}[!ht]
\centering
  \subfloat{}{\includegraphics[width=0.35\textwidth]{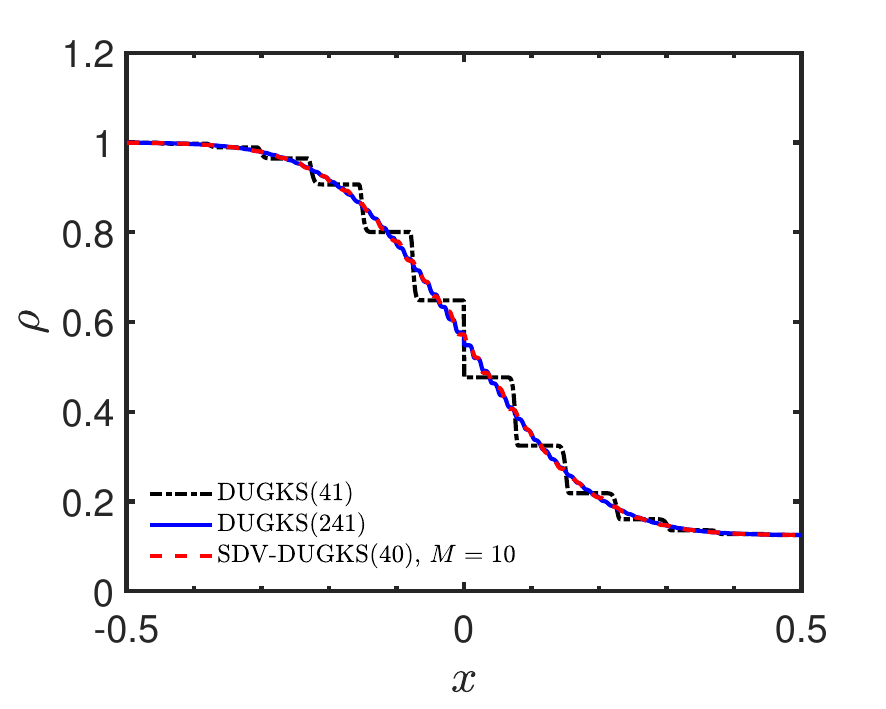}}~~
      \hspace{2mm}
  \subfloat{}{\includegraphics[width=0.35\textwidth]{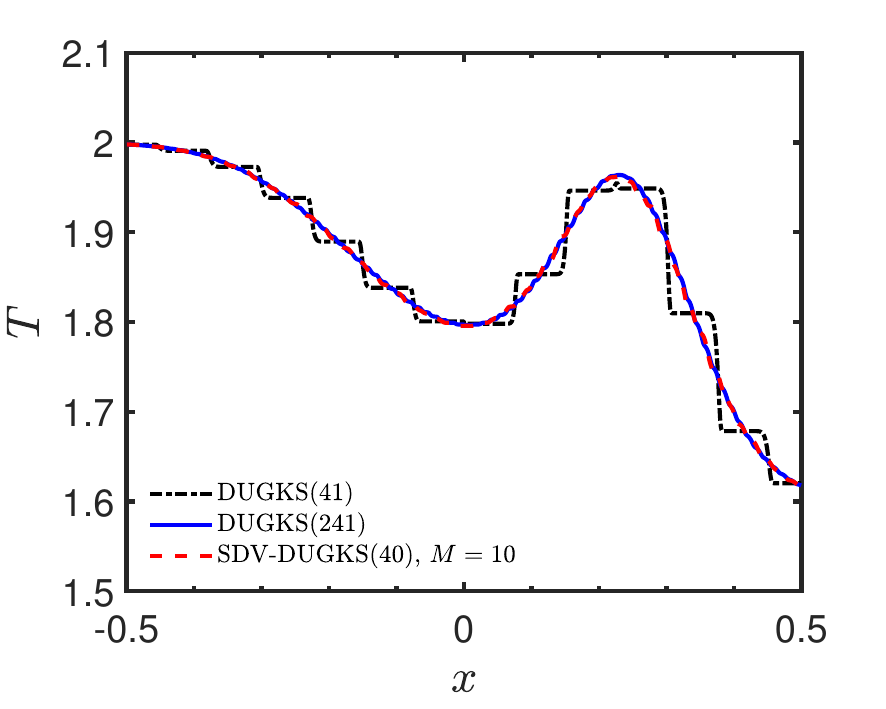}}~~
      \hspace{2mm}
  \subfloat{}{\includegraphics[width=0.35\textwidth]{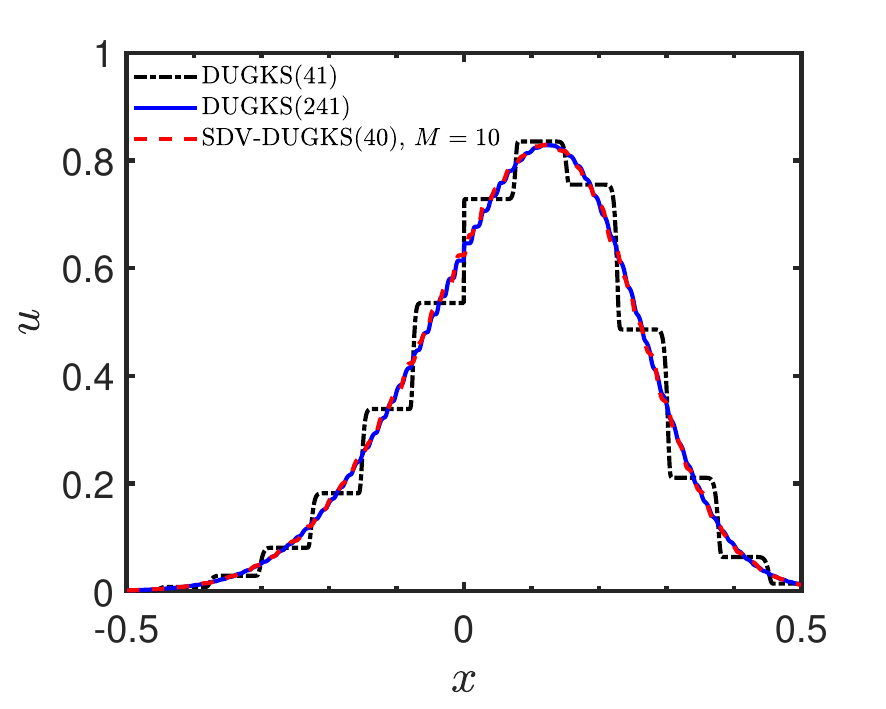}}~~
 \caption{Density, temperature, and velocity profiles of the Sod shock tube problem with $\mathrm{Kn} \to \infty$.}
\label{Sod_shock_tube411}
\end{figure}

\begin{figure}[!ht]
\centering
  \subfloat{}{\includegraphics[width=0.35\textwidth]{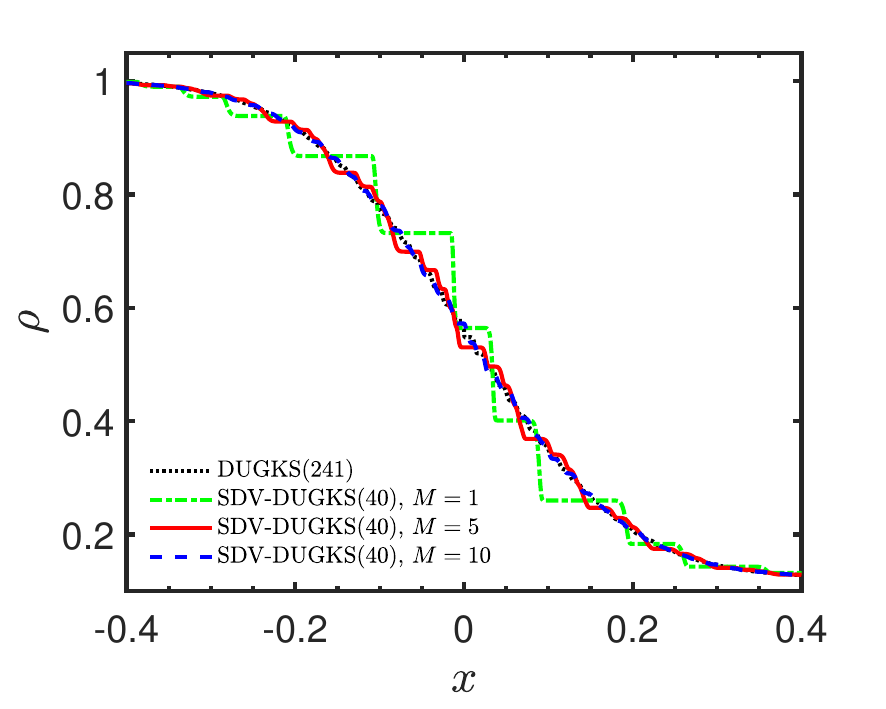}}~~
      \hspace{2mm}
  \subfloat{}{\includegraphics[width=0.35\textwidth]{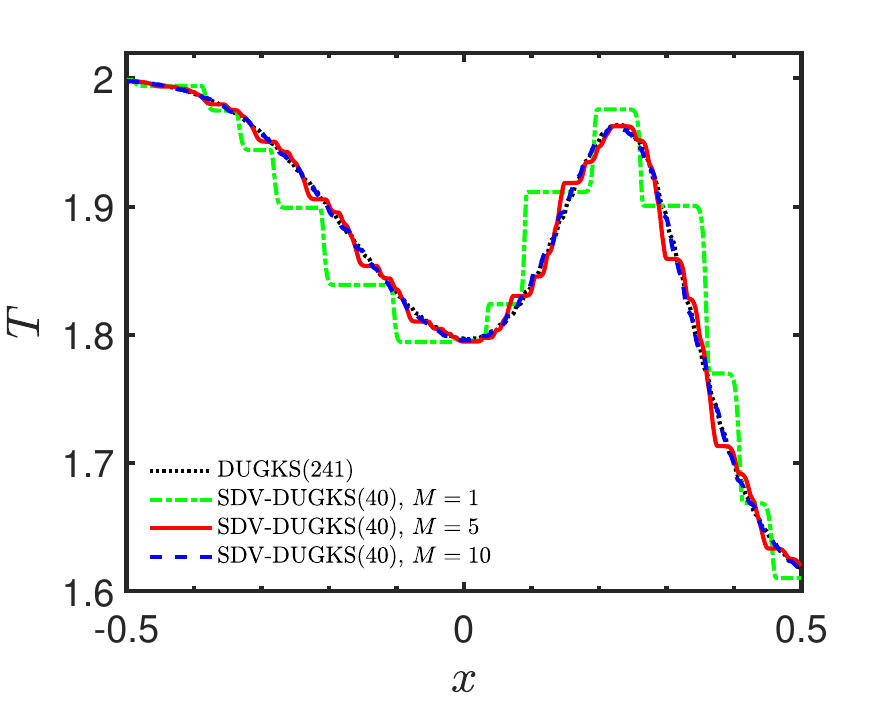}}~~
      \hspace{2mm}
  \subfloat{}{\includegraphics[width=0.35\textwidth]{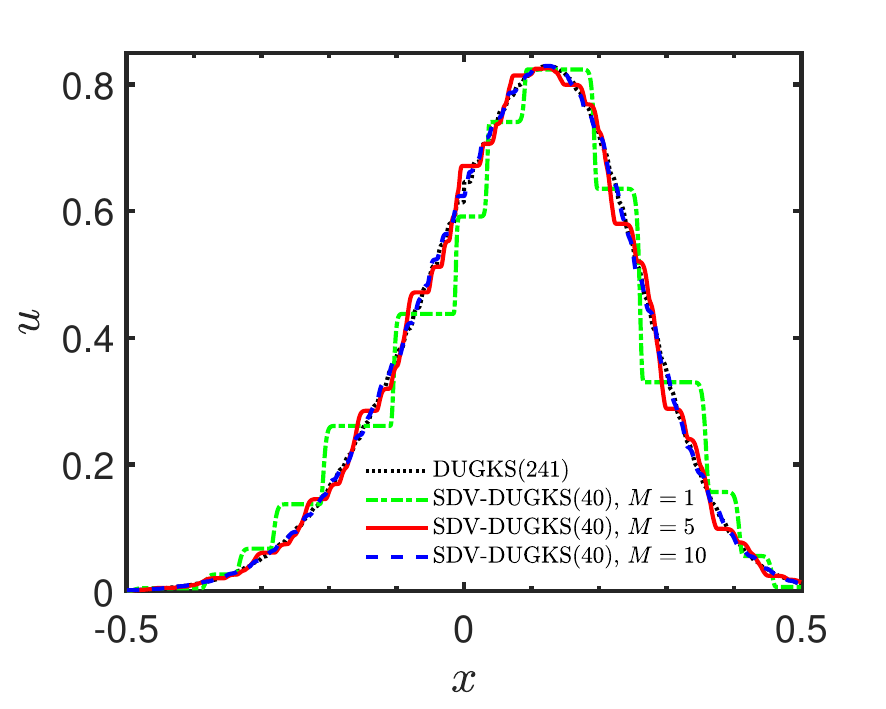}}~~
 \caption{SDV-DUGKS results with different numbers of realizations $M$ for the Sod shock tube problem with $\mathrm{Kn} \to \infty$.}
\label{averaging412}
\end{figure}

In this subsection, the standard Sod shock tube problem~\cite{sod1978survey} with an initial discontinuity is considered, which is ideal for assessing the capability of the present method in mitigating ray effects at highly rarefied conditions where the Knudsen number $\mathrm{Kn} \to \infty$.
The initial condition is
\begin{equation}
 (\rho,u,T)=\left\{
    \begin{array}{l l}
	{{(\rho_{1},u_{1},p_{1})=(1,\ 0,\ 1),}}            & {{x < 0,}}          \\
	{{(\rho_{2},u_{2},p_{2})=(0.125,\ 0,\ 0.1),}}      & {{x > 0.}}          \\
   \end{array}\right.
   \label{eq:shocktubeProblemIC}
\end{equation}
To make the ray effect clearly visible, the physical domain $[-0.5, 0.5]$ is divided into $1000$ uniformly distributed cells.
The time step is set to $\Delta{t}=5 \times 10^{-5}$, and the output time is $t_e = 0.15$.
To effectively reduce ray effects, the original DUGKS discretizes the velocity space domain $[-10, 10]$ into $241$ discrete velocity points based on the trapezoidal quadrature rule~\cite{rahman1990characterization}.
Meanwhile, the present SDV-DUGKS samples $40$ velocity points within the same velocity space domain.
Each realization is advanced to the output time $t_e$, and the results presented are averaged over $M = 10$ independent realizations.
Fig.~\ref{Sod_shock_tube411} presents the density, temperature, and velocity profiles at $t_e$ obtained by both methods, together with the DUGKS result using $40$ uniform velocity space grids.
As shown, the macroscopic profiles computed with DUGKS using $40$ uniform velocity grids exhibit pronounced ray effects.
In contrast, the result from SDV-DUGKS using $40$ sampled velocity points per realization agrees well with the DUGKS result using $240$ uniform velocity grids, and neither shows significant ray effects.
For this test case, SDV-DUGKS requires only about $1/3$ of the total memory (peak resident set size (RSS): $8924$ KB compared to $28016$ KB) compared to DUGKS while achieving comparable results.

In addition, Fig.~\ref{averaging412} compares the macroscopic profiles obtained by SDV-DUGKS with different numbers of realizations, using the DUGKS result with $240$ uniform velocity grids as reference.
The SDV-DUGKS result, obtained by the proposed averaging strategy over $M$ realizations, effectively mitigates ray effects as $M$ increases, and eventually agrees with the reference data in that no significant ray effects are present.

\subsection{$\it{1}$D Riemann problem}\label{sec32}
\begin{figure}[!ht]
\centering
  \subfloat{}{\includegraphics[width=0.35\textwidth]{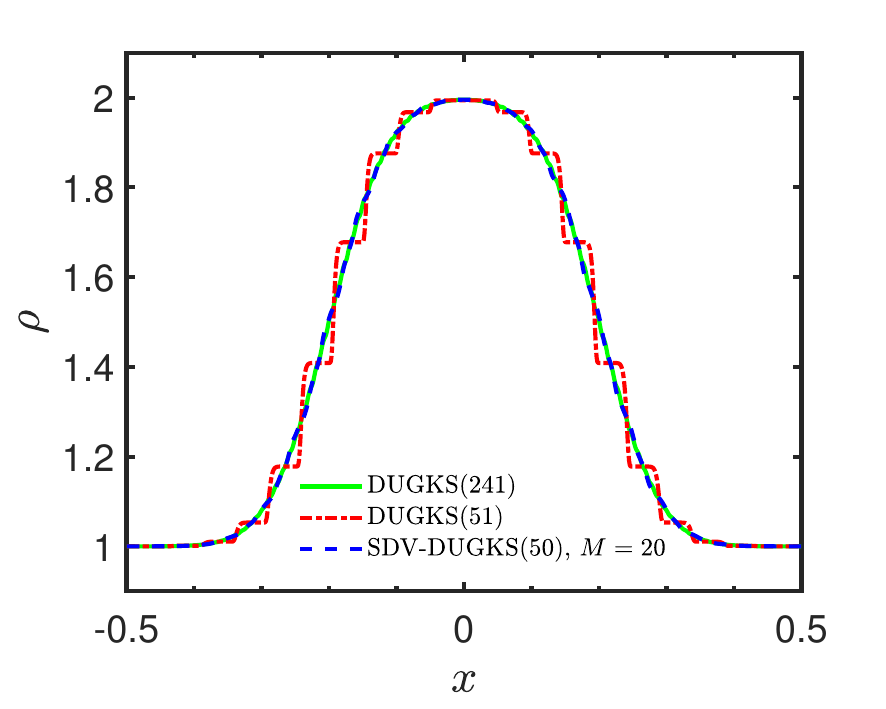}}~~
      \hspace{2mm}
  \subfloat{}{\includegraphics[width=0.35\textwidth]{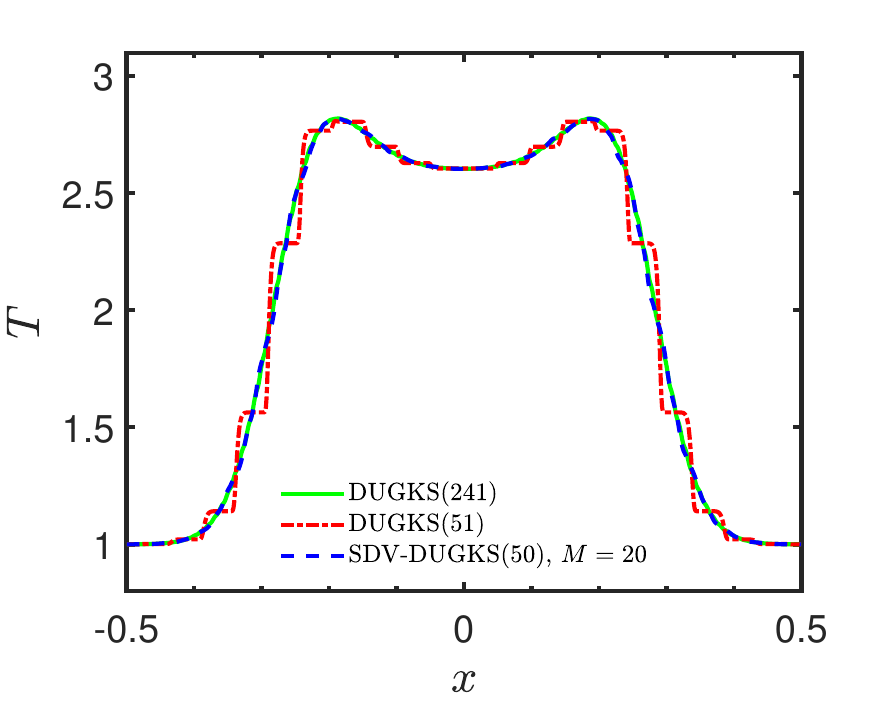}}~~
      \hspace{2mm}
  \subfloat{}{\includegraphics[width=0.35\textwidth]{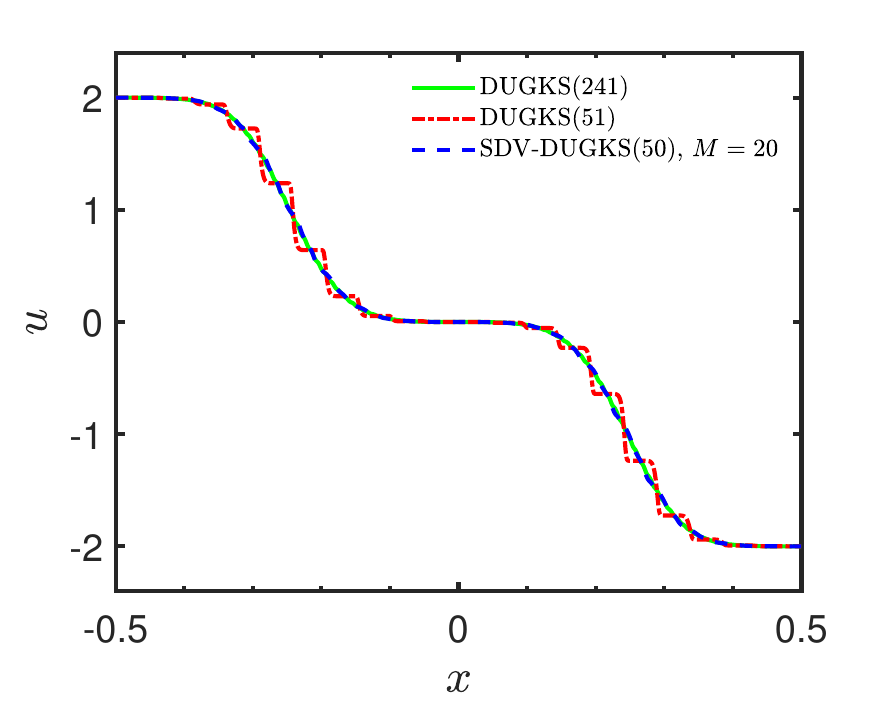}}~~
 \caption{Density, temperature, and velocity profiles of the $1\mathrm{D}$ Riemann problem with $\mathrm{Kn} \to \infty$.}
\label{Riemann_problem421}
\end{figure}

\begin{figure}[!ht]
\centering
  \subfloat{}{\includegraphics[width=0.35\textwidth]{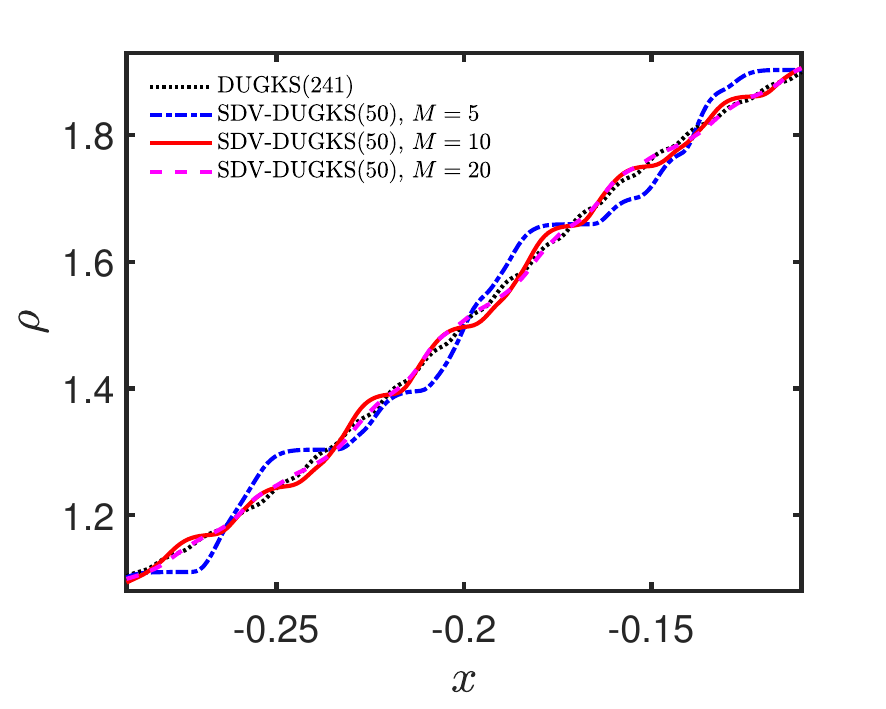}}~~
      \hspace{2mm}
  \subfloat{}{\includegraphics[width=0.35\textwidth]{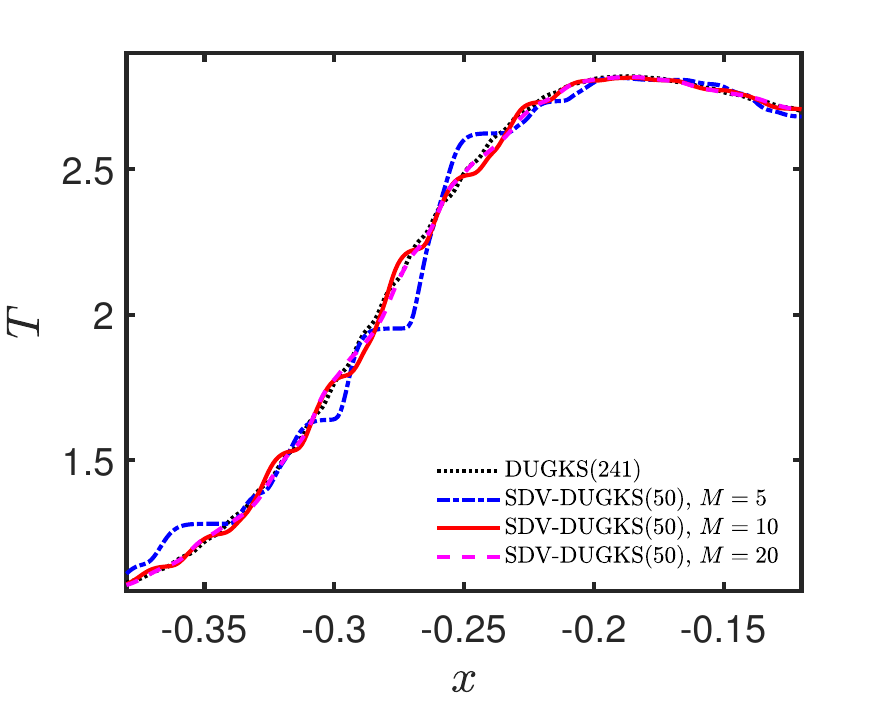}}~~
      \hspace{2mm}
  \subfloat{}{\includegraphics[width=0.35\textwidth]{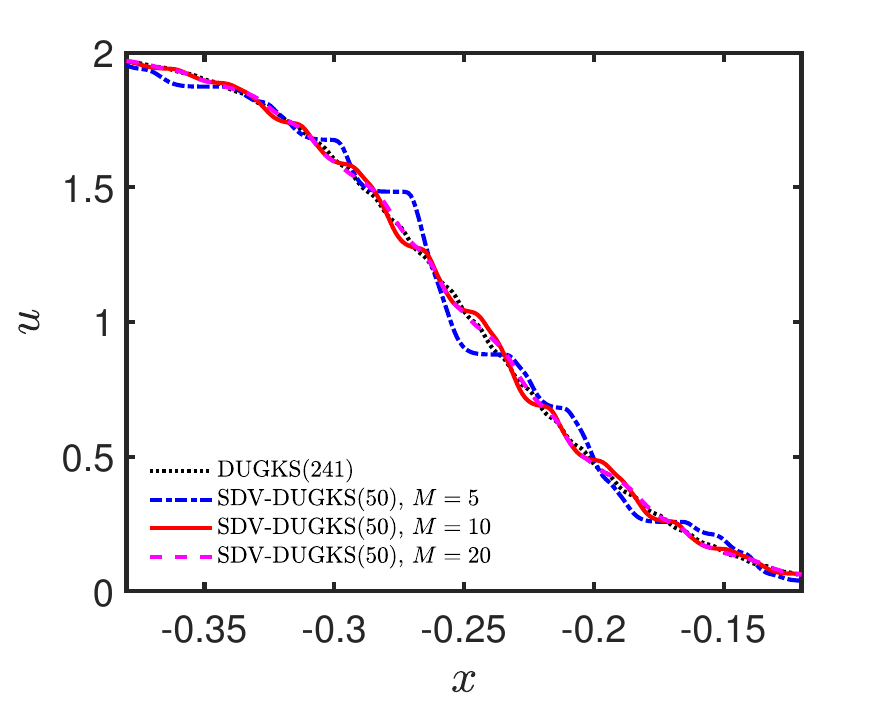}}~~
 \caption{SDV-DUGKS results with different numbers of realizations $M$ for the $1\mathrm{D}$ Riemann problem with $\mathrm{Kn} \to \infty$.}
\label{averaging422}
\end{figure}

Next, a $1\mathrm{D}$ Riemann problem involving the collision of two uniform streams is simulated under the condition $\mathrm{Kn} \to \infty$ to validate SDV-DUGKS.
The same configuration as in Ref.~\cite{zhu2020ray} (a modified version of the classical configuration from Ref.~\cite{toro2013riemann}) is adopted, with the initial condition given by
\begin{equation}
 (\rho,u,T)=\left\{
    \begin{array}{l l}
	{{(\rho_{1},u_{1},p_{1})=(1,\ 2,\ 1),}}            & {{x < 0,}}          \\
	{{(\rho_{2},u_{2},p_{2})=(1,\ -2,\ 1),}}      & {{x > 0.}}          \\
   \end{array}\right.
   \label{eq:shocktubeProblemIC}
\end{equation}
For this case, the same numerical setup as in Section~\ref{sec31} is adopted.
Specifically, the physical domain $[-0.5, 0.5]$ is divided into $1000$ uniformly distributed cells, and the time step is set to $\Delta t = 5 \times 10^{-5}$.
To significantly mitigate ray effects, the original DUGKS discretizes the velocity space domain $[-12, 12]$ into $241$ discrete velocity points based on the trapezoidal quadrature rule.
In contrast, SDV-DUGKS employs $50$ velocity points sampled from the same velocity space domain.
Each realization is advanced to the output time $t_e$, and the presented results are obtained by averaging over $M = 20$ independent realizations.
Figure~\ref{Riemann_problem421} presents the macroscopic profiles at $t_e = 0.1$ obtained by both methods.
To highlight the improvement by SDV-DUGKS, the DUGKS result using $50$ uniform velocity grids is included, which exhibits clearly visible ray effects.
By comparison, the SDV-DUGKS result using $50$ sampled velocity points per realization closely matches the DUGKS result with $241$ deterministic velocity points, with neither exhibiting noticeable ray effects.
Moreover, SDV-DUGKS consumes only about $1/3$ of the total memory (peak RSS: $10028$ KB compared to $27996$ KB) relative to DUGKS, yet delivers comparable results.

In addition, Figure~\ref{averaging422} compares the macroscopic profiles obtained by SDV-DUGKS with different numbers of realizations, against the DUGKS result with $241$ deterministic velocity points.
The ray effects in the SDV-DUGKS result become progressively less observable as the number of realizations $M$ increases.
At $M = 20$, the SDV-DUGKS result agrees well with the DUGKS reference and exhibits ray effects of similar magnitude.

\subsection{$\it{2}$D lid-driven cavity flow}\label{sec33}
\begin{figure}[!ht]
\centering
  \subfloat{\includegraphics[width=0.25\textwidth]{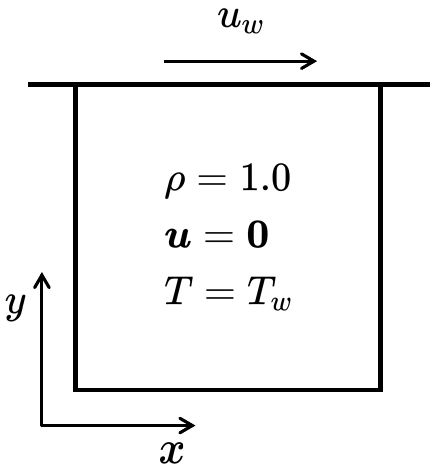}}~~
 \caption{The schematic of the $2\mathrm{D}$ lid-driven cavity flow.}
\label{2D_cavity}
\end{figure}

The $2\mathrm{D}$ lid-driven cavity flow~\cite{kuhlmann2018lid} is a fundamental benchmark problem.
When simulated with deterministic methods that discretize the velocity space, the discontinuity of the distribution function in velocity space at the boundary inherently gives rise to the ray effect~\cite{sone1992discontinuity,zhu2020ray}.
To demonstrate the capability of SDV-DUGKS, we consider this classic problem at the collisionless limit $\mathrm{Kn} \to \infty$.
The computational domain, as shown in Fig.~\ref{2D_cavity}, is a square region with side length $L = 1$.
The lid moves in the positive $x$-direction with velocity $u_w = 0.274$, while all walls are maintained at a constant temperature $T_w = 1.099$.
Diffuse reflection boundary conditions~\cite{guo2013discrete,li2005application} applied to all boundaries.
Initially, the fluid is at rest with density $\rho_0 = 1.0$ and temperature $T_0 = T_w$.
The computational domain $[0,1]^2$ is discretized using a uniform Cartesian grid of $60 \times 60$ cells for both the original DUGKS and the present SDV-DUGKS.
The time step $ \Delta t $ is determined by the Courant-Friedrichs-Lewy (CFL) condition~\cite{de2013courant}:
\begin{equation}
   \Delta t=\beta\frac{\Delta{x}}{|\bm\xi|_{\operatorname*{max}}},
\label{eq:CFLnumber}
\end{equation}
where $\beta = 0.5$ is the CFL number for this case.

\begin{figure}[!ht]
\centering
  \subfloat[]{\includegraphics[width=0.35\textwidth]{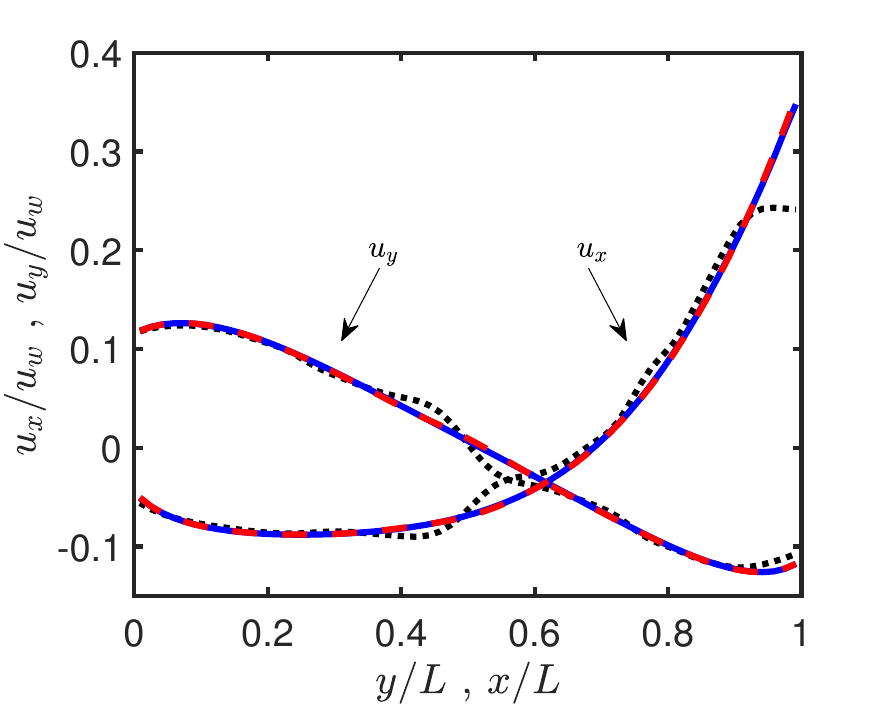}}~~
      \hspace{4mm}
  \subfloat[]{\includegraphics[width=0.35\textwidth]{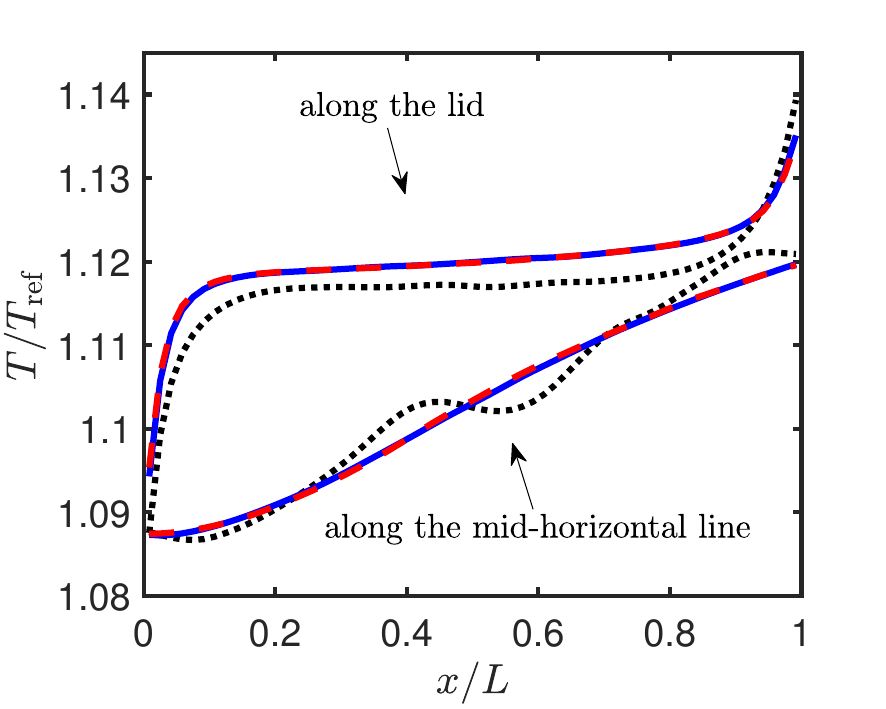}}~~
 \caption{Velocity and temperature profiles along the selected lines for the $2\mathrm{D}$ lid-driven cavity flow with $\mathrm{Kn} \to \infty$. (a) $u_x$-velocity profile along the mid-vertical line and $u_y$-velocity profile along the mid-horizontal line; (b) temperature profile along the mid-horizontal line and the lid. Blue solid line: Reference solution; black dotted line: DUGKS results; red dashed line: SDV-DUGKS results.}
\label{MicrocavityFlow_u&T_432}
\end{figure}

\begin{figure}[!ht]
\centering
  \subfloat{\includegraphics[width=0.9\textwidth]{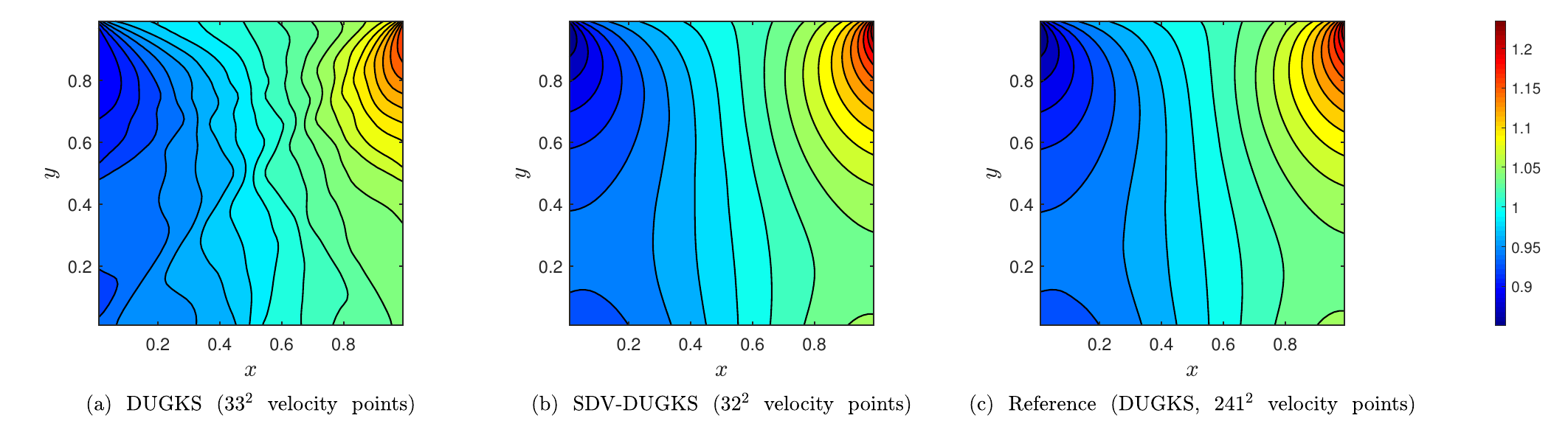}}~~
 \caption{The density contour for the $2\mathrm{D}$ lid-driven cavity flow with $\mathrm{Kn} \to \infty$. (a) DUGKS results ($33^2$ velocity points); (b) SDV-DUGKS results ($32^2$ velocity points); (c) Reference data (DUGKS, $241^2$ velocity points).}
\label{2D_cavity_rhocontour_433}
\end{figure}

\begin{figure}[!ht]
\centering
  \subfloat{\includegraphics[width=0.9\textwidth]{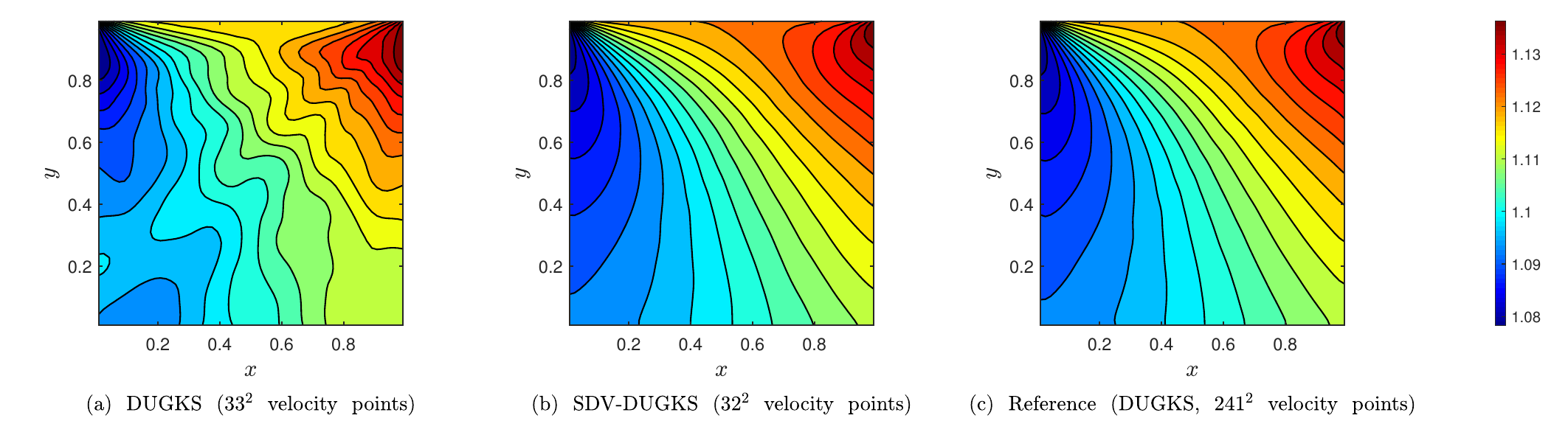}}~~
 \caption{The temperature contour for the $2\mathrm{D}$ lid-driven cavity flow with $\mathrm{Kn} \to \infty$. (a) DUGKS results ($33^2$ velocity points); (b) SDV-DUGKS results ($32^2$ velocity points); (c) Reference data (DUGKS, $241^2$ velocity points).}
\label{2D_cavity_Tcontour_434}
\end{figure}

\begin{figure}[!ht]
\centering
  \subfloat{\includegraphics[width=1.055\textwidth]{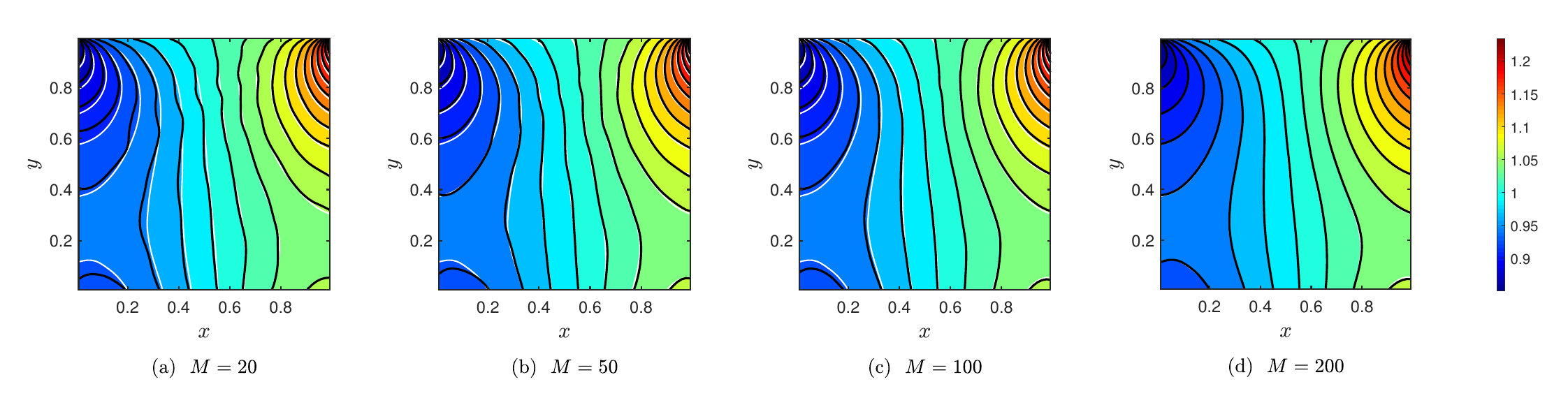}}~~
 \caption{The density contour for the $2\mathrm{D}$ lid-driven cavity flow with $\mathrm{Kn} \to \infty$. White solid lines with the colored background: Reference data; black solid lines: SDV-DUGKS results. (a) $M = 20$; (b) $M = 50$; (c) $M = 100$; (d) $M = 200$.}
\label{2D_cavity_rhocontour_MC_435}
\end{figure}

\begin{figure}[!ht]
\centering
  \subfloat{\includegraphics[width=1.055\textwidth]{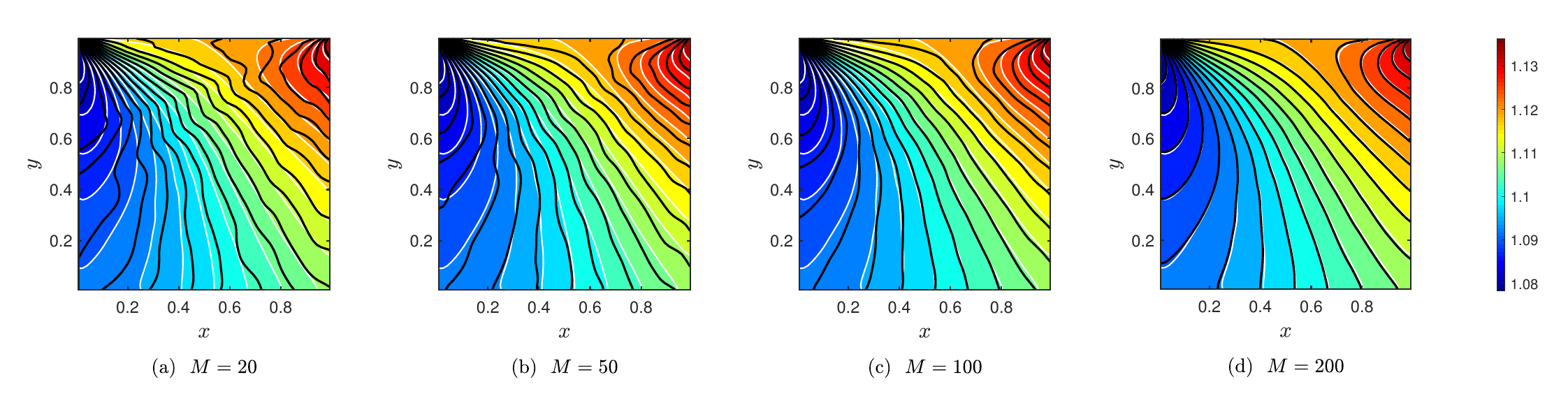}}~~
 \caption{The temperature contour for the $2\mathrm{D}$ lid-driven cavity flow with $\mathrm{Kn} \to \infty$. White solid lines with the colored background: Reference data; black solid lines: SDV-DUGKS results. (a) $M = 20$; (b) $M = 50$; (c) $M = 100$; (d) $M = 200$.}
\label{2D_cavity_Tcontour_MC_436}
\end{figure}

The original DUGKS generates two results based on the trapezoidal quadrature rule: one employs a uniform discretization of the velocity space $[-6, 6]^2$ with $33^2$ points, and the other uses a finer discretization with $241^2$ points serving as the reference solution.
The convergence criterion for DUGKS is defined as
\begin{equation}
\varepsilon_{k} = \frac{ \sum_{j=1}^{N_c} | W_{j,k}^{n+1} - W_{j,k}^n | }{ \sum_{j=1}^{N_c} | W_{j,k}^n | } < 1.0 \times 10^{-8}, \quad \forall \, k \in \{1, 2, \dots, D+2\},
\label{eq:convergence_criterion}
\end{equation}
where $k$ denotes the $k$-th component of $\bm{W}_j = (\rho_j, \rho_j \bm{u}_j, \rho_j E_j)^T$, and $N_c$ is the total number of cells.
For SDV-DUGKS, $32^2$ velocity points are randomly sampled from a uniform $32 \times 32$ stratification within the same velocity space domain.
A looser convergence criterion of $1.0 \times 10^{-6}$ is used for each realization:
\begin{equation}
\varepsilon^{\prime}_{k} = \frac{ \sum_{j=1}^{N_c} | W_{j,k}^{n+1} - W_{j,k}^n | }{ \sum_{j=1}^{N_c} | W_{j,k}^n | } < 1.0 \times 10^{-6}, \quad \forall \, k \in \{1, 2, \dots, D+2\}.
\label{eq:convergence_criterion_MC}
\end{equation}
Averages are then performed over $M = 200$ realizations for this case.
Fig.~\ref{MicrocavityFlow_u&T_432} presents the velocity and temperature profiles along the selected lines, comparing the DUGKS results with $33^2$ velocity points, the SDV-DUGKS results with $32^2$ sampled velocity points, and the reference solution. Figs.~\ref{2D_cavity_rhocontour_433} and~\ref{2D_cavity_Tcontour_434} show the density and temperature contours, respectively, for the same three results.
Under comparable total memory usage (peak RSS: 558232 KB for SDV-DUGKS compared to 593152 KB for DUGKS), the SDV-DUGKS results exhibit no visible ray effects, whereas ray effects are clearly observed in the DUGKS results.
Furthermore, despite requiring only about $1/55$ of the memory (peak RSS: 558232 KB compared to 30524512 KB), the SDV-DUGKS results agree well with the reference solution.

Figs.~\ref{2D_cavity_rhocontour_MC_435} and~\ref{2D_cavity_Tcontour_MC_436} present the density and temperature contours obtained by SDV-DUGKS with different numbers of realizations $M$, together with the reference data.
As $M$ increases, more velocity points are sampled across realizations, allowing distribution functions at an increasing number of velocity points to contribute to the final averaged result, thereby progressively reducing ray effects.
This trend is clearly observed from the figures.
Notably, due to the stochastic nature of sampling, not every individual realization, when averaged, improves the accuracy of the final result.
Nevertheless, the overall trend is toward improved accuracy as more realizations are included.

\subsection{$\it{2}$D Riemann problem}\label{sec34}
\begin{figure}[!ht]
\centering
  \subfloat{\includegraphics[width=0.9\textwidth]{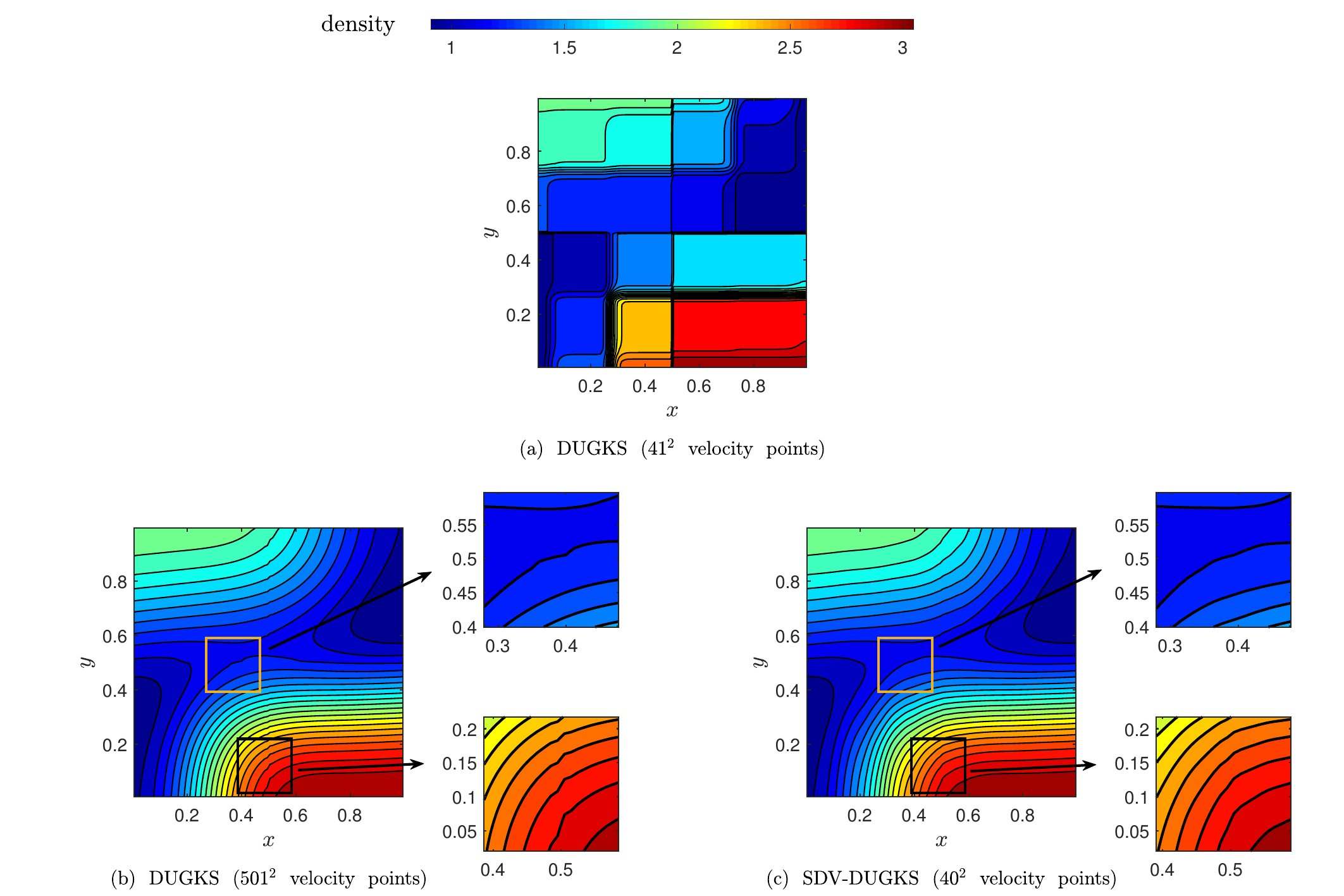}}~~
 \caption{The density contour for the $2\mathrm{D}$ Riemann problem with $\mathrm{Kn} \to \infty$. (a) DUGKS results ($41^2$ velocity points); (b) SDV-DUGKS results ($40^2$ velocity points); (c) DUGKS results ($501^2$ velocity points).}
\label{2D_Riemann_rhocontour_441}
\end{figure}

\begin{figure}[!ht]
\centering
  \subfloat{\includegraphics[width=0.9\textwidth]{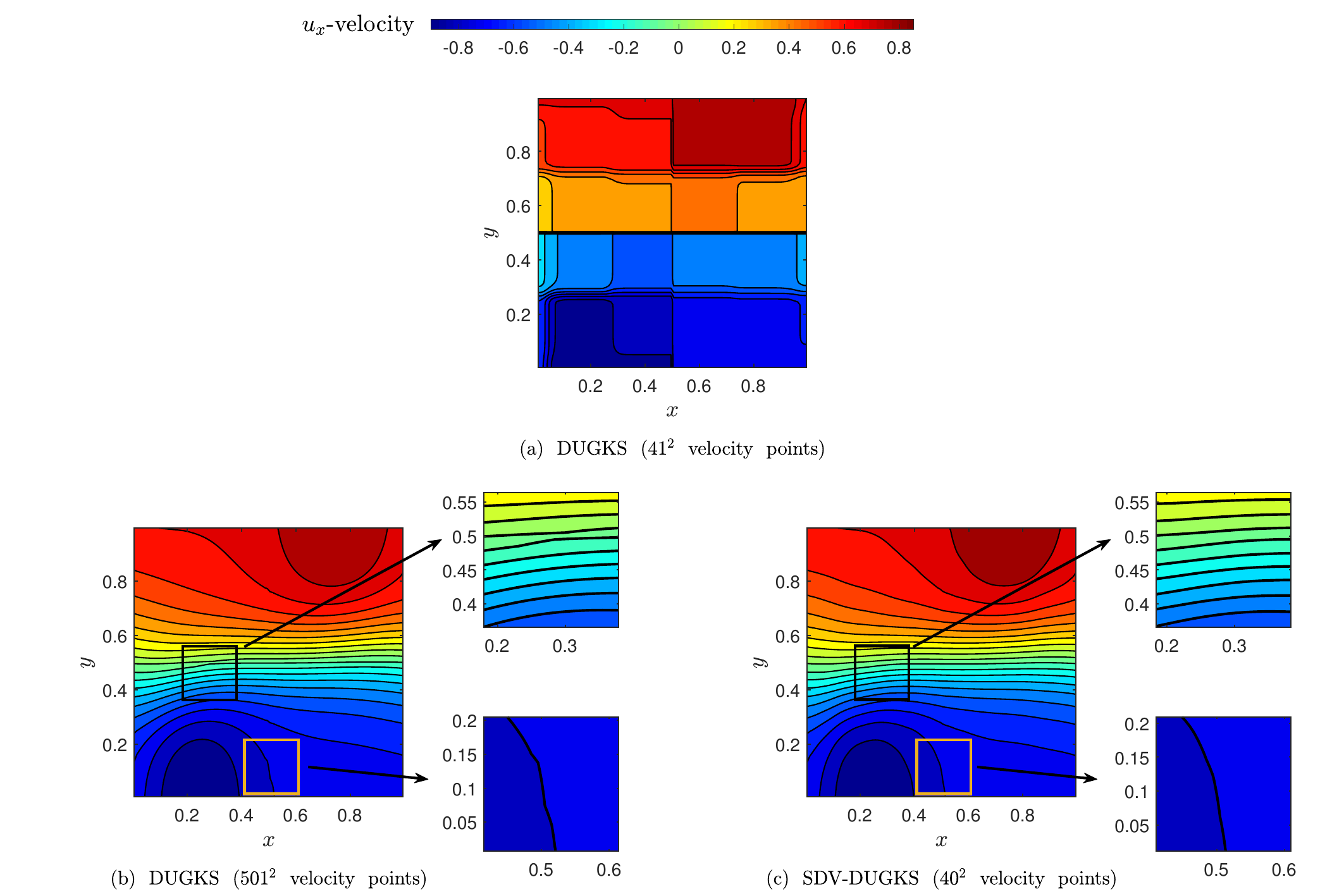}}~~
 \caption{The $u_x$-velocity contour for the $2\mathrm{D}$ Riemann problem with $\mathrm{Kn} \to \infty$. (a) DUGKS results ($41^2$ velocity points); (b) SDV-DUGKS results ($40^2$ velocity points); (c) DUGKS results ($501^2$ velocity points).}
\label{2D_Riemann_ucontour_442}
\end{figure}

\begin{figure}[!ht]
\centering
  \subfloat{\includegraphics[width=0.9\textwidth]{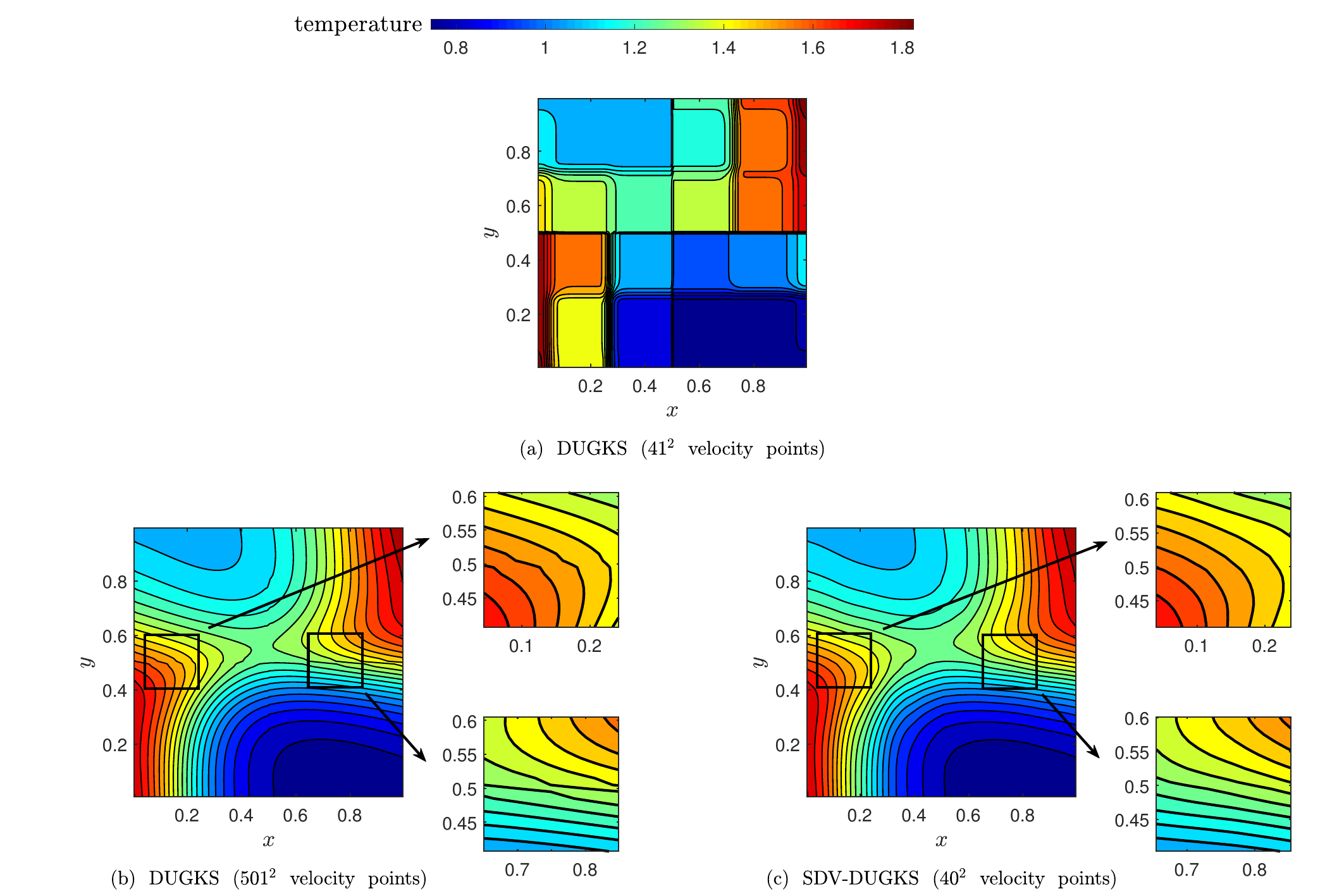}}~~
 \caption{The temperature contour for the $2\mathrm{D}$ Riemann problem with $\mathrm{Kn} \to \infty$. (a) DUGKS results ($41^2$ velocity points); (b) SDV-DUGKS results ($40^2$ velocity points); (c) DUGKS results ($501^2$ velocity points).}
\label{2D_Riemann_rhocontour_443}
\end{figure}

In this subsection, the $2\mathrm{D}$ Riemann problem with initial discontinuities is simulated to validate SDV-DUGKS for $2\mathrm{D}$ transient cases in the collisionless limit $\mathrm{Kn} \to \infty$.
A classical configuration from Ref.~\cite{lax1998solution} is adopted, with the initial condition given by
\begin{equation}
 (\rho,u,v,p)=\left\{
    \begin{array}{l l}
	{{(\rho_{1},u_{1},v_{1},p_{1})=(1,\ 0.75,\ -0.5,\ 1),}}                 & {{x > 0.5,\quad y > 0.5,}}               \\
	{{(\rho_{2},u_{2},v_{2},p_{2})=(2,\,0.75,\,0.5,\,1),}}                  & {{x < 0.5,\quad y > 0.5,}}            \\
	{{(\rho_{3},u_{3},v_{3},p_{3})=(1,\,-0.75,\,0.5,\,1),}}                 & {{x < 0.5,\quad y < 0.5,}}         \\
	{{(\rho_{4},u_{4},v_{4},p_{4})=(3,\,-0.75,\,-0.5,\,1),}}                & {{x > 0.5,\quad y < 0.5.}}            \\
   \end{array}\right.
   \label{eq:RiemannProblemIC}
\end{equation}
The specific heat ratio is set to $\gamma = 1.4$.
No-flux boundary conditions are applied to all four boundaries.
In this simulation, a uniform Cartesian grid of $100 \times 100$ cells is employed to discretize the computational domain $[0, 1]^2$, for both the original DUGKS and the present SDV-DUGKS.
The time step is set to $\Delta t = 4 \times 10^{-4}$, and the output time is $t_e = 0.3$.

For the original DUGKS, two uniform discretizations of the velocity space $[-15, 15]^2$ are considered based on the trapezoidal quadrature rule: one using $41^2$ points and the other $501^2$ points.
For the present SDV-DUGKS, $40^2$ velocity points are randomly sampled within the same velocity space domain via a uniform $40 \times 40$ stratification.
Each realization is simulated until $t_e$, after which the final results are obtained by averaging over $M = 200$ independent realizations.
Figs.~\ref{2D_Riemann_rhocontour_441}, \ref{2D_Riemann_ucontour_442}, and \ref{2D_Riemann_rhocontour_443} present the contours of density, $u_x$-velocity, and temperature, respectively, for the three results.
In the DUGKS results with $41^2$ velocity points, marked ray effects are observed, characterized by regular plateau-like structures.
When the velocity points are increased to $501^2$, these plateau-like structures are largely eliminated, although a few faint jagged structures still remain, indicating the presence of weak ray effects.
Despite requiring only about $1/157$ of the memory (peak RSS: 2304880 KB compared to 361009624 KB), the SDV-DUGKS results exhibit even weaker ray effects, showing only fainter jagged structures.

\begin{figure}[!ht]
\centering
  \subfloat{\includegraphics[width=1.055\textwidth]{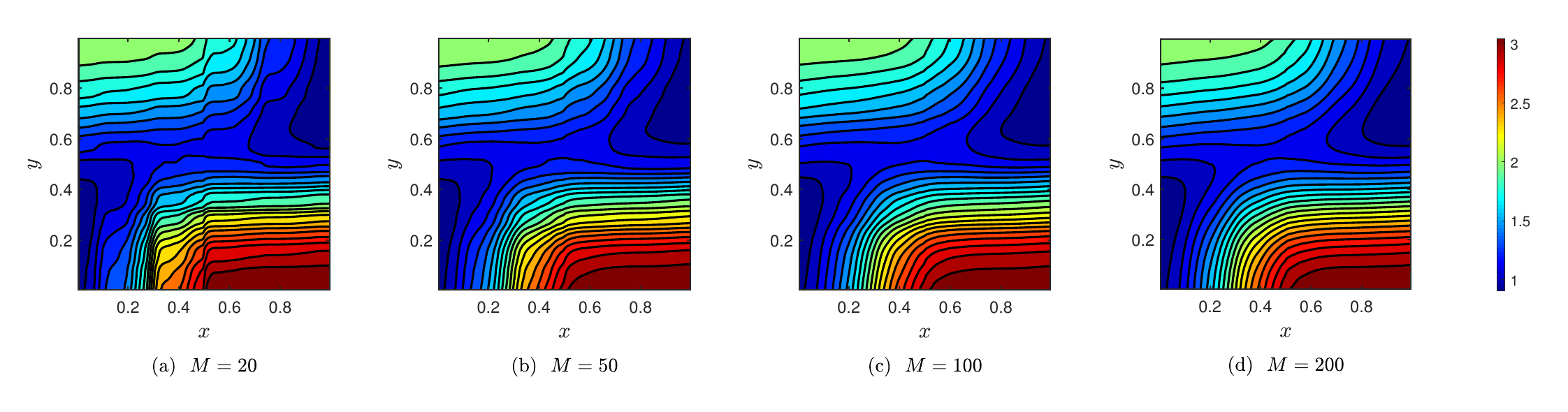}}~~
 \caption{The density contour for the $2\mathrm{D}$ Riemann problem with $\mathrm{Kn} \to \infty$. Black solid lines with the colored background: SDV-DUGKS results. (a) $M = 20$; (b) $M = 50$; (c) $M = 100$; (d) $M = 200$.}
\label{2D_Riemann_rhocontour_MC_444}
\end{figure}

\begin{figure}[!ht]
\centering
  \subfloat{\includegraphics[width=1.055\textwidth]{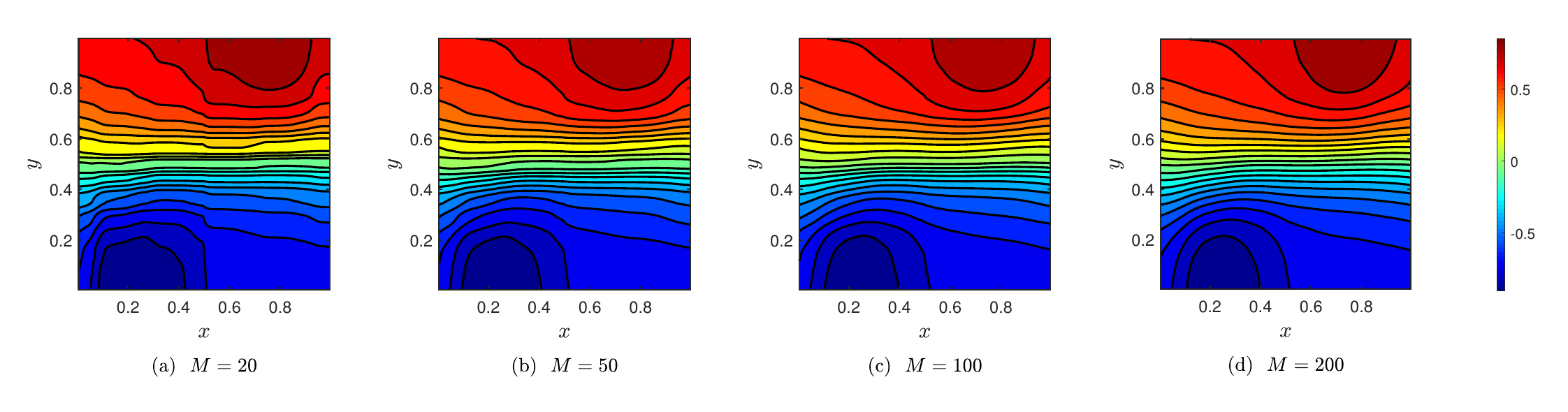}}~~
 \caption{The $u_x$-velocity contour for the $2\mathrm{D}$ Riemann problem with $\mathrm{Kn} \to \infty$. Black solid lines with the colored background: SDV-DUGKS results. (a) $M = 20$; (b) $M = 50$; (c) $M = 100$; (d) $M = 200$.}
\label{2D_Riemann_ucontour_MC_445}
\end{figure}

\begin{figure}[!ht]
\centering
  \subfloat{\includegraphics[width=1.055\textwidth]{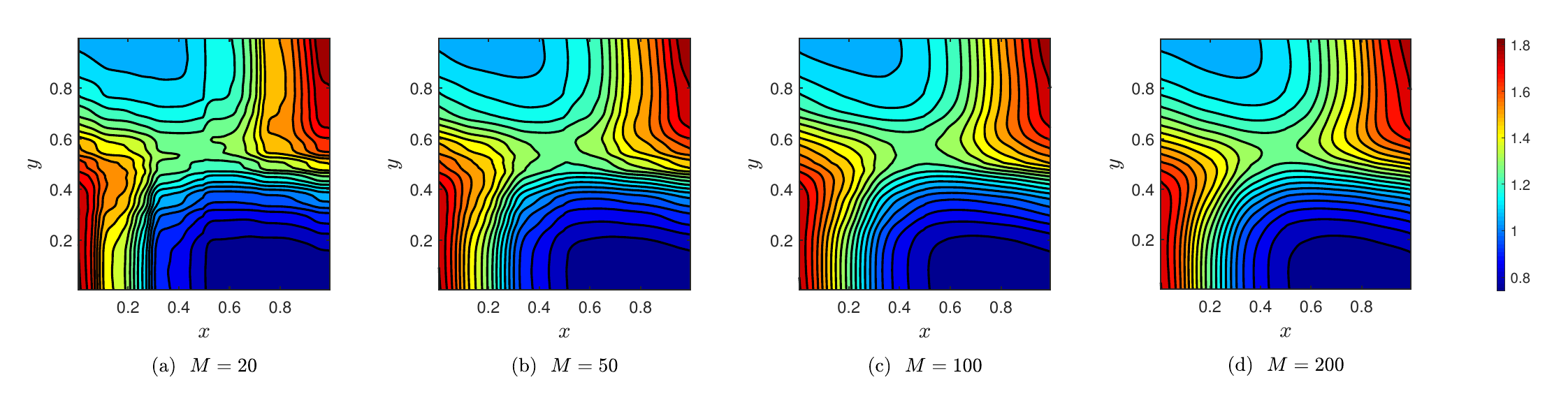}}~~
 \caption{The temperature contour for the $2\mathrm{D}$ Riemann problem with $\mathrm{Kn} \to \infty$. Black solid lines with the colored background: SDV-DUGKS results. (a) $M = 20$; (b) $M = 50$; (c) $M = 100$; (d) $M = 200$.}
\label{2D_Riemann_Tcontour_MC_446}
\end{figure}

Figures \ref{2D_Riemann_rhocontour_MC_444}, \ref{2D_Riemann_ucontour_MC_445}, and \ref{2D_Riemann_Tcontour_MC_446} present the density, temperature, and $u_x$-velocity contours, respectively, obtained using the SDV-DUGKS method with different numbers of realizations $M$.
As $M$ increases, the ray effects weaken, and the results evolve from exhibiting large-scale plateau-like structures, to showing weakened jagged structures, and finally to retaining only faint jagged structures.

\section{Conclusion}\label{sec4}
This study extends an ensemble-of-subproblems strategy with stochastic discrete velocities to deterministic methods to mitigate ray effects in rarefied flow simulations.
We incorporate this strategy within the DUGKS framework, and the resulting method is denoted as SDV-DUGKS.
The strategy involves performing multiple independent simulations, each using a small set of randomly sampled velocity points, and then averaging the solutions to obtain the final result.

The effectiveness of SDV-DUGKS in mitigating ray effects is assessed through a comparative study with the original DUGKS based on four test cases: the Sod shock tube problem, the $1$D Riemann problem, the $2$D lid-driven cavity flow, and the $2$D Riemann problem.
Based on the numerical results in this work, the main findings are summarized as follows:
\begin{enumerate}[(1)]
\item For one-dimensional cases in the collisionless limit $\mathrm{Kn} \to \infty$, SDV-DUGKS reduces memory usage by approximately $2/3$ compared with that of the original DUGKS while achieving good agreement;
\item For two-dimensional cases in the collisionless limit $\mathrm{Kn} \to \infty$, SDV-DUGKS requires one to two orders of magnitude less memory than the original DUGKS while achieving good agreement;
\item SDV-DUGKS is effective in mitigating ray effects.
\end{enumerate}

In the current implementation, the stratified sampling employs a stratification based on a structured velocity space grid.
However, it may not be universally efficient for all problems.
Further improvements in sampling efficiency are the focus of our future work.

\section*{Acknowledgments}
Zhaoli Guo acknowledges the support provided by the National Natural Science Foundation of China (grant no.12472290).
Weidong Li is grateful to the support provided by National Key Laboratory of Aerospace Physics in Fluids (grant no.KT-APF-2024-004).


\bibliographystyle{elsarticle-num-names}
\bibliography{ray-effects}

\end{document}